\documentclass[apj,twocolumn,revtex4,twocolappendix]{emulateapj}
\usepackage{graphicx}
\usepackage{bm}
\usepackage[dvips]{color}
\usepackage{color}
\usepackage{xspace}
\usepackage{ulem}
\usepackage{lscape}
\usepackage{natbib}
\usepackage{multirow}
\usepackage{booktabs}
\usepackage{anyfontsize}
\usepackage{adjustbox}

\newcommand{\swift}{\textit{Swift}\xspace}
\newcommand{\nustar}{\textit{NuSTAR}\xspace}
\newcommand{\chandra}{\textit{Chandra}\xspace}

\newcommand{\myarcsec}{\hbox{$.\!\!^{\prime\prime}$}}
\newcommand{\ergs}{erg\,s$^{-1}$}

\newcommand{\ergcms}{erg cm$^{-2}$ s$^{-1}$}
\DeclareRobustCommand{\ion}[2]{\textup{#1\,\textsc{\lowercase{#2}}}}

\shorttitle{Powerful AGN and outflow histories in low-$z$ LABs}
\shortauthors{
T. Kawamuro, M. Schirmer et al.
}

\begin{document}

\title{
  \nustar hard x-ray data and Gemini 3D spectra reveal 
  powerful AGN and outflow histories in two low-redshift Lyman-$\alpha$ blobs
}

\author{
 Taiki Kawamuro\altaffilmark{1,2},
 Mischa Schirmer\altaffilmark{3},
 James E. H. Turner\altaffilmark{3},
 Rebecca L. Davies\altaffilmark{4}, and
 Kohei Ichikawa\altaffilmark{2,5,6}
}

\altaffiltext{1}{Department of Astronomy, Kyoto University, Kyoto 606-8502, Japan}
\altaffiltext{2}{National Astronomical Observatory of Japan, Osawa, Mitaka, Tokyo 181-8588, Japan}
\altaffiltext{3}{Gemini Observatory, Southern Operations Center, Casilla 603, La Serena, Chile}
\altaffiltext{4}{Max-Planck-Institut f\"ur Extraterrestrische Physik, 85748 Garching, Germany}
\altaffiltext{5}{Department of Physics and Astronomy, University of Texas at San Antonio, One UTSA Circle, San Antonio, TX 78249, USA}
\altaffiltext{6}{Department of Astronomy, Columbia University, 550 West 120th Street, New York, NY 10027, USA}

\begin{abstract}
We have shown that Lyman-$\alpha$ blobs (LABs) may still exist even at $z\sim0.3$, about 7 billion years later than most other LABs known \citep{Sch16}. Their luminous Ly$\alpha$ and [\ion{O}{III}] emitters at $z\sim0.3$ offer new insights into the ionization mechanism. This paper focuses on the two X-ray brightest LABs at $z\sim0.3$, SDSS J0113+0106 (J0113) and SDSS J1155$-$0147 (J1155), comparable in size and luminosity to `B1', one of the best-studied LABs at $z \gtrsim$ 2. Our \nustar hard X-ray (3--30 keV) observations reveal powerful active galactic nuclei (AGN) with $L_{2-10{\;\rm keV}}=(0.5$--$3)\times10^{44}$ \ergs. J0113 also faded by a factor of $\sim 5$ between 2014 and 2016, emphasizing that variable AGN may cause apparent ionization deficits in LABs. Joint spectral analyses including \chandra data constrain column densities of $N_{\rm H}=5.1^{+3.1}_{-3.3}\times10^{23}$ cm$^{-2}$ (J0113) and $N_{\rm H}=6.0^{+1.4}_{-1.1}\times10^{22}$ cm$^{-2}$ (J1155). J0113 is likely buried in a torus with a narrow ionization cone, but ionizing radiation is also leaking in other directions as revealed by our Gemini/GMOS 3D spectroscopy. The latter shows a bipolar outflow over $10$ kpc, with a peculiar velocity profile that is best explained by AGN flickering. X-ray analysis of J1155 reveals a weakly absorbed AGN that may ionize over a wide solid angle, consistent with our 3D spectra. Extinction corrected [\ion{O}{III}] log-luminosities are high, $\sim43.6$. The velocity dispersions are low, $\sim100$--$150$ km s$^{-1}$, even at the AGN positions. We argue that this is a combination of high extinction hiding the turbulent gas, and previous outflows that have cleared the escape paths for their successors.

\end{abstract}

\bibliographystyle{apj}


\section{INTRODUCTION}\label{sec:int}
Lyman-$\alpha$ blobs (LABs) are extended ($\sim$ 20--200 kpc) nebulae with luminosities of
$L_{{\rm Ly}\alpha}=10^{42-44}$ \ergs. They are commonly found at redshifts $z\gtrsim2$ in
overdense regions and proto-clusters \citep{Mat11,Erb11}, and have been associated with a
broad range of galaxies hosting active galactic nuclei (AGN), star-burst sub-mm galaxies, 
and passively evolving red galaxies \citep[e.g.][]{Fra01,Mat04,Gea09,Bri13}. LABs are landmarks 
of massive galaxy formation \citep{Gea05,Mat06,Pre08}, and this understanding would greatly 
benefit from studying the physical conditions and ionizing sources in LABs.

This, however, is difficult for several reasons. First, Ly$\alpha$ photons scatter resonantly 
in space and frequency \citep{Mei02,Ver06,Kol10}, and non-resonant diagnostic lines such as [\ion{O}{III}] 
and H$\alpha$ must be used to study the physical conditions in the gas. Second, any emission lines 
in high-$z$ LABs are faint because of severe cosmological surface brightness dimming
($\propto (1+z)^{-4}$). 
Third, the lines may not be observable from the ground if their redshifted emission falls into the 
near-infrared atmospheric absorption bands. Fourth, the fraction of Ly$\alpha$ photons that manage to 
escape is controlled by dust, neutral hydrogen, metallicity, and 
outflows \citep{Cia14,Hen15,Riv15,Yan16}, a process that is still poorly understood.

Many LABs lack evidence of ionizing continuum sources that would explain their high Ly$\alpha$ luminosities.
Three main ideas have been put forward to solve this mystery:

First, \textit{the ionizing sources are spatially distributed}, i.e., Ly$\alpha$ 
is created by other processes, such as shock heating by starburst driven superwinds 
\citep[e.g.][]{Tan00}, or gravitational cooling radiation of neutral hydrogen
\citep[`cold accretion',][]{Hai00,Far01,Nil06,Ros12}.

Second, \textit{the ionizing sources are hidden}, in particular obscured powerful AGN \citep{Bas04,Gea09,Ove13,Pre15}. It is difficult to distinguish between buried AGN and
cold accretion \citep{Pre15}, and between cold accretion and scattering \citep{Lau07,Tre16},
because the same data can be interpreted differently. Also, \cite{Ste00} and \cite{Sch16}
caution against postulating the presence of \textit{obscured} AGN when evidence for other mechanisms is absent.

Third, \textit{the ionizing sources (AGN) are variable}, and thus not detectable at
all times \citep{Sch16}. Accordingly, during their quasar phases, the AGN fill the
LABs with Ly$\alpha$ photons; the latter are trapped because of resonant scattering
and released only gradually. The observable Ly$\alpha$ light curve is thus a very damped
response to the AGN's ionizing light curve \citep{Roy10,Xu11}. Effectively, the
Ly$\alpha$ and X-ray light curves are decorrelated. High Ly$\alpha$ luminosities therefore
do not require \textit{currently} powerful AGN \citep{Sch16}, because AGN spend most of the time 
in a low accretion state \citep{Nov11}. This may explain many of the non-detections
of AGN that were originally attributed to heavy obscuration.

The existence of LABs at low redshift would provide exciting targets because of high 
 fluxes and the full availability of optical diagnostic emission lines.
Previous searches
for LABs at $z\sim0.7$--$1.1$ using {\it GALEX} grism spectroscopy were unsuccessful
\citep{Kee09,Wol14}, apart from one detection by \cite{Bar12} of a less luminous LAB
($L_{\rm Ly\alpha}=7.2\times10^{42}$ \ergs) at $z=0.977$. These were first strong indications
that LABs must rapidly disappear from the Universe between $z\sim2$ and $z\sim0.7$ \cite[see Figure 6 in][]{Sch16}.
\cite{Ove13} have predicted that a small number of low-$z$ LABs should still exist in
low-density environments, mostly powered by AGN because cold accretion streams would have
ceased. Indeed, in \cite{Sch16} we have reported the discovery of rare, luminous LABs at
$z\sim0.3$ by cross-correlating the most powerful [\ion{O}{III}] emitters of \cite{Sch13}
with archival \textit{GALEX} far-UV images. \textit{Hubble Space Telescope} ({\it HST}) 
far-UV images and spectra have been scheduled (PI: Schirmer, proposal ID \#14749) to ultimately 
verify the Ly$\alpha$ luminosities.

As predicted in Section~4.2 of \cite{Ove13}, \cite{Sch13} confirmed that these low-$z$ LABs 
are indeed powered by AGN from their high [\ion{O}{III}]/H$\beta$ line ratios. If a duty cycle 
of a luminous AGN phase is short as predicted by \cite{Nov11}, the majority of these AGN is likely to be weak. 
This idea corresponds to the third one raised previously. Hence, for the examination, we have 
analyzed the \chandra data, and found that the soft ($\lesssim8$\,keV) X-ray luminosities are lower 
than expected \citep{Sch16}. This apparently favors our hypothesis.
However, due to the soft energy band and the poor photon statistics, the discussion highly depends on 
the assumed absorption correction. The AGN could simply be heavily obscured ($N_{\rm H}\gtrsim 10^{24}$ 
cm$^{-2}$) rather than intrinsically weak. This has motivated us to conduct further X-ray observations 
at high energies to constrain the column densities and the \textit{current} luminosities. 

\begin{deluxetable*}{lccccccccccccccccc}
\tablecaption{Basic target information \label{tab:info_srcs}}
\tablewidth{0pt}
\tablehead{
Name & $\alpha_{\rm J2000}$ & $\delta_{\rm J2000}$ & $z$ & $D$ & Observatory & ID & Date & Exp. \\ 
(1)  & (2) & (3) & (4) & (5)  & (6) & (7) & (8) & (9) 
}
\startdata
\object{SDSS J011341.11+010608.5}  & 18.42129  & 1.10237  & 0.281 & 1.45 & {\it NuSTAR}    & 60201054002   & 2016 July 01 & 21 \\
        &           &          &       &      & {\it Chandra}   & 16102         & 2014 Jun 02 & 15 \\ 
        &           &          &       &      & {\it Swift}     & 00081905001   & 2016 Jun 30 & 1.6 \\ 
        &           &          &       &      & Gemini/GMOS     & GS-2016B-DD-4 & 2016 Nov 06 & 4.8 \\  \hline 
\object{SDSS J115544.59$-$014739.9}  & 178.93580 & -1.79443 & 0.306 & 1.60 & {\it NuSTAR}    & 60201055002   & 2016 Jun 07 & 20 \\
        &           &          &       &      & {\it Chandra}   & 3140          & 2002 Dec 02 & 30 \\
        &           &          &       &      & {\it Swift}     & 00081906001   & 2016 Jun 05 & 1.9 \\ 
        &           &          &       &      & Gemini/GMOS     & GS-2014B-Q-63 & 2015 Mar 31 & 5.4 
\enddata
\tablecomments{
Columns: 
(1) Galaxy name. 
(2)-(3) Right Ascension and Declination in units of degree. 
(4) Redshift. 
(5) Luminosity distance in units of Gpc based on each redshift. 
(6) Observatory 
(7)-(9) 
Details for each observation 
(Obs. ID, Obs. date, and total exposure time in units of ksec). 
}
\end{deluxetable*}

\nustar is the most sensitive observatory to hard X-rays ($>10$ keV), which are less affected
by obscuration. Strong constraints on the luminosity can be obtained even for Compton-thick
AGN ($N_{\rm H} > 10^{24}$ cm$^{-2}$). Soft X-ray spectra of such AGN may be dominated by scattered
and/or reflected components, and thus a joint spectral analysis of soft and hard X-ray data is
required to accurately determine obscuration and luminosities. Using a pilot \nustar program,
we observed the two \chandra brightest (and least likely faded) sources of \cite{Sch16}, 
SDSS J011341.11+010608.5 and SDSS J115544.59$-$014739.9 (hereafter J0113 and J1155). 

This paper is organized as follows. Section~\ref{sec:obs} presents an overview of our 
\nustar, \chandra, \swift, and Gemini observations. Details about the simultaneous analysis 
of the good quality \nustar and \chandra spectra and its results are found in Section~\ref{sec:broad_ana}. 
Results from the optical 3D spectra 
are presented in Section~\ref{sec:Ana3Dspec}. In Section~\ref{sec:results}, we discuss the
structures of the AGN, and the morphologies and kinematics of the extended emission line
regions. 
We summarize and conclude in Section~\ref{sec:sum}. Throughout this paper, we assume
a $\Lambda$CDM cosmology with $H_0=70$ km\,s$^{-1}$ Mpc$^{-1}$, $\Omega_{\rm m}=0.3$ and
$\Omega_{\Lambda}=0.7$. Unless otherwise noted, all errors concerning the X-ray data
are quoted at the 90\% confidence level for a single parameter of interest; errors for the
optical data are 67\% confidence level.

\section{OBSERVATIONS AND DATA REDUCTION}\label{sec:obs}
 
\begin{figure*}[!ht]
\begin{center}\vspace{.8cm}
\includegraphics[scale=0.3,angle=-90]{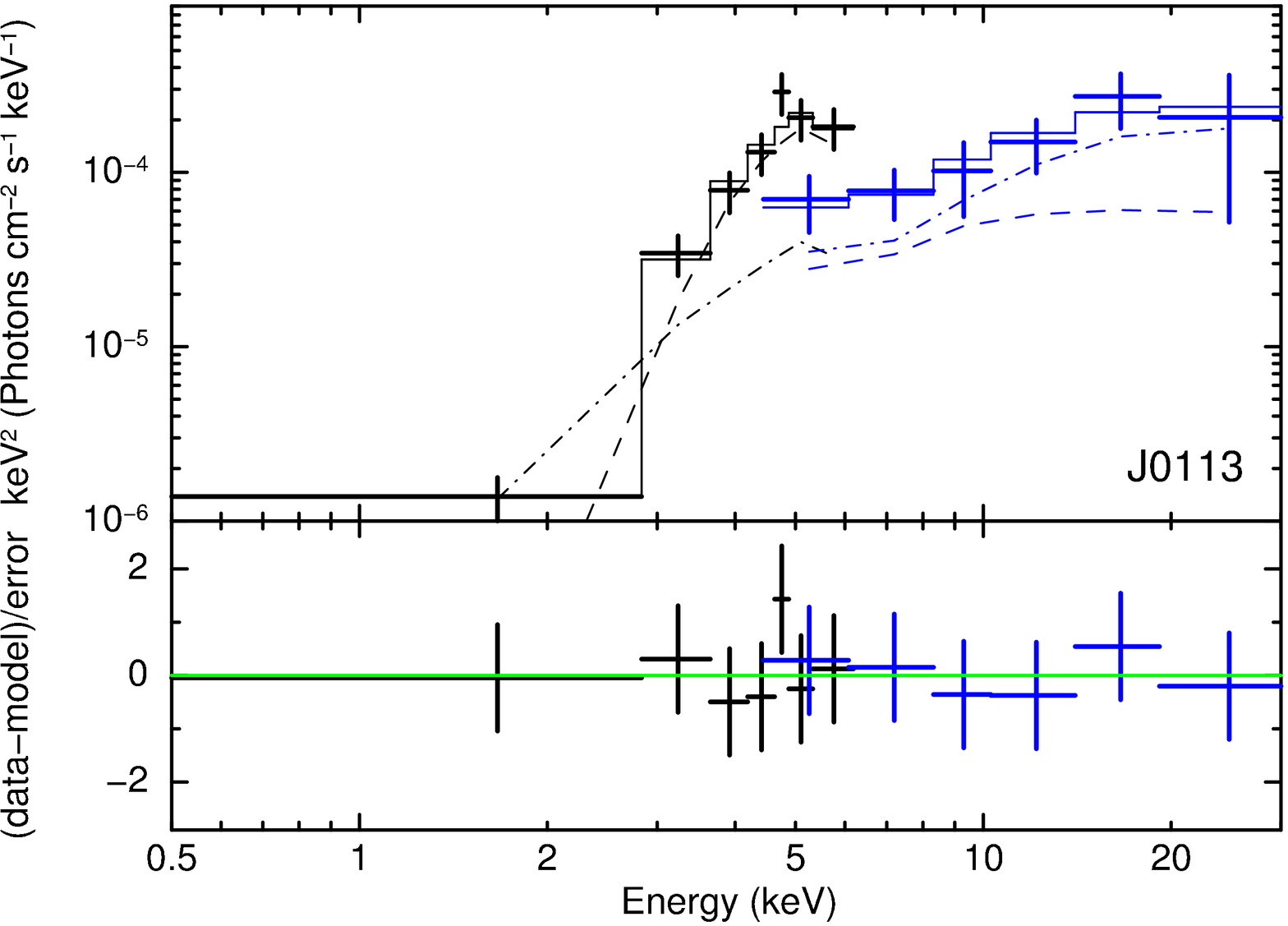} \hspace{1cm}
\includegraphics[scale=0.3,angle=-90]{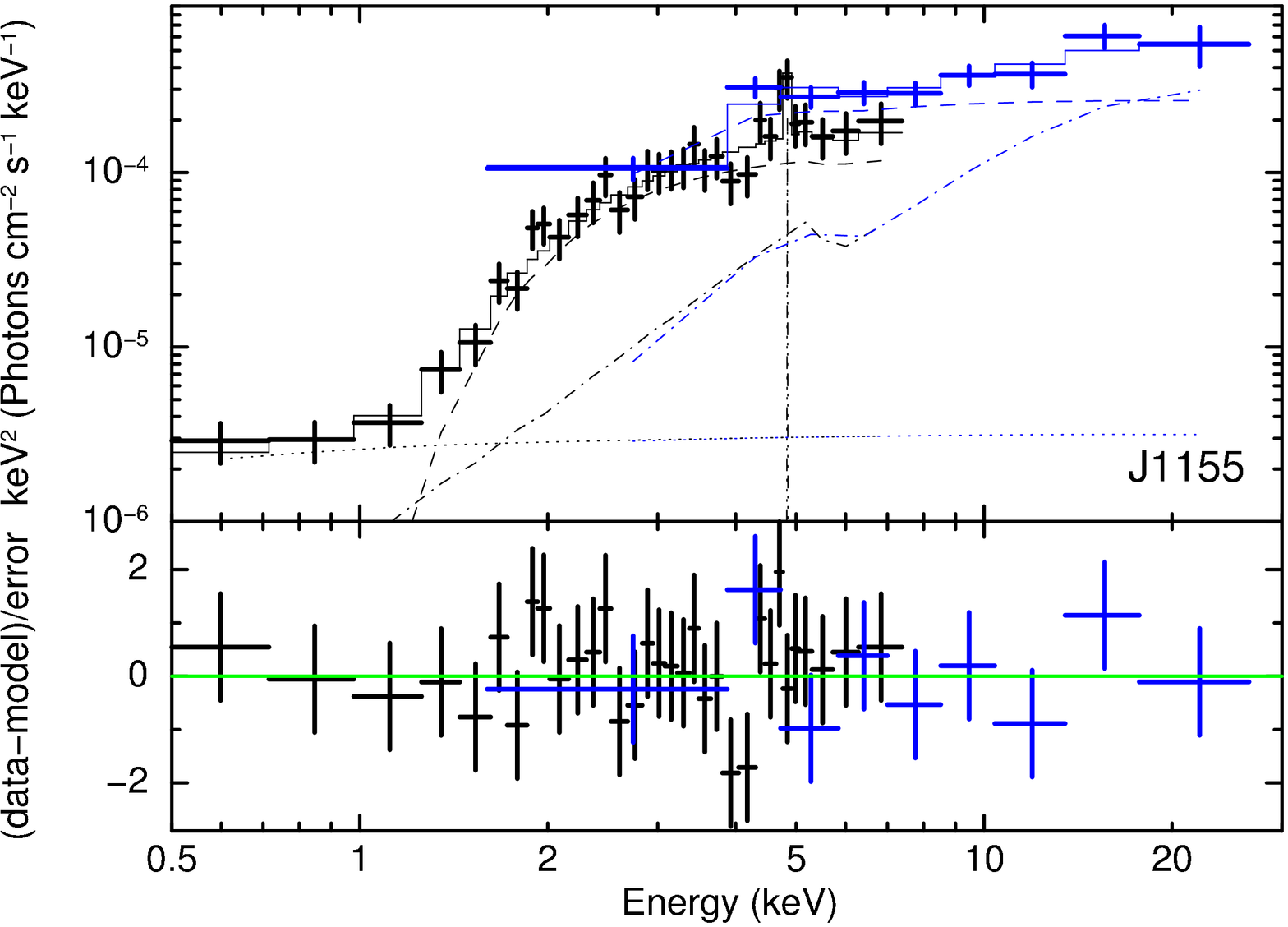} \vspace{-.8cm}
\caption{{\label{fig:src_spec} 
Unfolded broad-band spectra of J0113 (left) and J1155 (right). The \chandra and \nustar data are displayed with thick lines 
in black and blue, respectively. Overlaid thin lines are the best-fit model (solid line), the absorbed cut-off power-law 
component (dashed), the reflection component (dot-dashed), the scattered component (dotted), and the iron K$\alpha$ 
line (dot-dot-dashed). In the left, the scattered and iron-K$\alpha$ components cannot be seen because of their small 
contributions.  The lower panels display the residuals between the data and the best-fit model.
}}
\end{center}
\end{figure*}

Table~\ref{tab:info_srcs} summarizes the basic information of the two LABs and the details on 
their \nustar \citep{Har13}, \chandra, and \swift observations. The \nustar observations were conducted 
in cycle 2 as a Guest Observatory program (PI: Schirmer), followed by the \swift 
observations with short exposures ($\sim$ 1--2 ksec) as a service program. Also, \chandra observed J0113 (PI: Schirmer) 
and J1155 (PI: Coppi) in 2014 and 2002, respectively. 
We use the \swift/X-ray telescope \cite[XRT;][]{Geh05} data reprocessed through 
the online tools provided by the UK \swift Science Data Centre \citep{Eva09}\footnote{{\tt http://www.swift.ac.uk/user\_objects/}}. 
Due to the low signal-to-noise ratio of the \swift spectra (see Section~\ref{sec:swift_xrt}), 
we perform the detailed analysis only for the \nustar and \chandra spectra (Figure~\ref{fig:src_spec}). 
XSPEC version 12.9.0.n \citep{Arn96} is used for the spectral fitting. 

The 3D Gemini/GMOS spectra are from a larger campaign targeting all 17 objects of
\cite{Sch16}; details are given below.

\subsection{\nustar Observations}

\nustar carries two independent focal plane modules (FPMA and FPMB), sensitive in the 3--80 keV range. We reprocess the 
event files using the standard {\tt nupipeline} script as described in the ``{\it NuSTAR} Analysis Quickstart Guide" 
\footnote{{\tt 
http://www.srl.caltech.edu/NuSTAR\_Public/NuSTAROperationSite\\/SAA\_Filtering/SAA\_Filter.php
}}.
Our targets are faint ($\sim10^{-3}$ cts\,s$^{-1}$), and periods of high background (such as passages through or near the South Atlantic Anomaly (SAA)) must be excluded. Typical background rates observed with \nustar are $\lesssim$ 1 cts s$^{-1}$, integrated over the focal plane \citep{For14}. Times of high background can be identified by simultaneously increased count rates in the detectors as well as in the shields that surround the focal planes. Using the telemetry reports made by the \nustar team, we first check the total event rates during all orbital passages of our observations. In the J1155 observation, the event rate increases around the standard SAA area (above $\sim $ 30 cts s$^{-1}$). We run {\tt nupipeline} with the option {\tt saamode=optimized} to reject times with high count rates in both the shields and detectors. High count rates may occasionally occur in the so-called tentacle region \citep{For14} near the SAA, and we exclude these times for J1155 by setting {\tt tentacle=yes}. Background rates during the observation of J0113 are stable and low ({\tt saamode=none} and {\tt tentacle=no}). 

The full width at half maximum (FWHM) of the \nustar point spread function is $\sim$ 20$^{\prime\prime}$. 
The source spectra are extracted from circular regions with a radius of $1^{\prime}$ centered on
the centroid of the counts. The background spectra are extracted from identical apertures, located in an 
off-source region on the same detector. 
To increase the signal-to-noise ratios, the FPMA and FPMB spectra are combined without 
renormalization due to lack of statistical power in the count rates using the {\tt addascaspec} 
command in {\tt FTOOLS}. J0113 and J1155 
are significantly detected in the 3--30 keV band with a signal-to-noise ratio of $\sim$ 8 and 20, respectively. 


\subsection{\chandra Observations}

The \chandra/ACIS data provide the soft energy band (0.5--10 keV) for the spectral analysis. 
The raw data are reprocessed by using the {\tt chandra\_repro} pipeline included in the \chandra 
Interactive Analysis of Observations (CIAO) software (version 4.8) and referring to the latest CALDB 
at the time (version 4.7.2). The typical angular resolution of \chandra/ACIS is $\sim1^{\prime\prime}$. 
Our targets are barely resolved in the \chandra data, and no offsets are observed between the
  X-ray and [\ion{O}{III}] peak emission \citep[see][for details]{Sch16}. The source spectra were
extracted
from a circular region with $2^{\prime\prime}$ radius, centered on the \chandra detections. The
background spectra are taken from similar off-source regions. Background correction is minimal, as 
the source count rates are $\sim100$ times higher. 
Consequently, we detect J0113 and J1155 at the 11$\sigma$ and 23$\sigma$ levels, respectively, 
in the 0.5--10 keV band.

\subsection{\swift Observations}\label{sec:swift_xrt} 

We analyze the \swift/XRT spectra covering the soft energy band (0.3--10 keV). We find the low 
detection significances of 2.3$\sigma$ and 3.6$\sigma$ for J0113 and J1155, respectively. It is likely 
due to the low exposure and effective area. Thus, these 
spectra are ignored in the detailed broadband X-ray spectral analysis. Here, we only constrain the 
observed fluxes to confirm the consistency with those estimated from the \nustar spectra. 
The powerlaw model is fitted based on the $C$-statistics appropriate to low-counts spectra 
instead of the chi-squared method. The 2--10 keV fluxes are found to be $< 2.0 \times 10^{-13}$ 
\ergcms for J0113 and $4.7^{+6.3}_{-2.9} \times 10^{-13}$ \ergcms for J1155. They are in well agreement 
with those from the \nustar spectra (see Section~\ref{sec:broad_ana} and Table~\ref{tab:para}).

\begin{figure*}[t]
\begin{center}
\includegraphics[width=1.0\hsize]{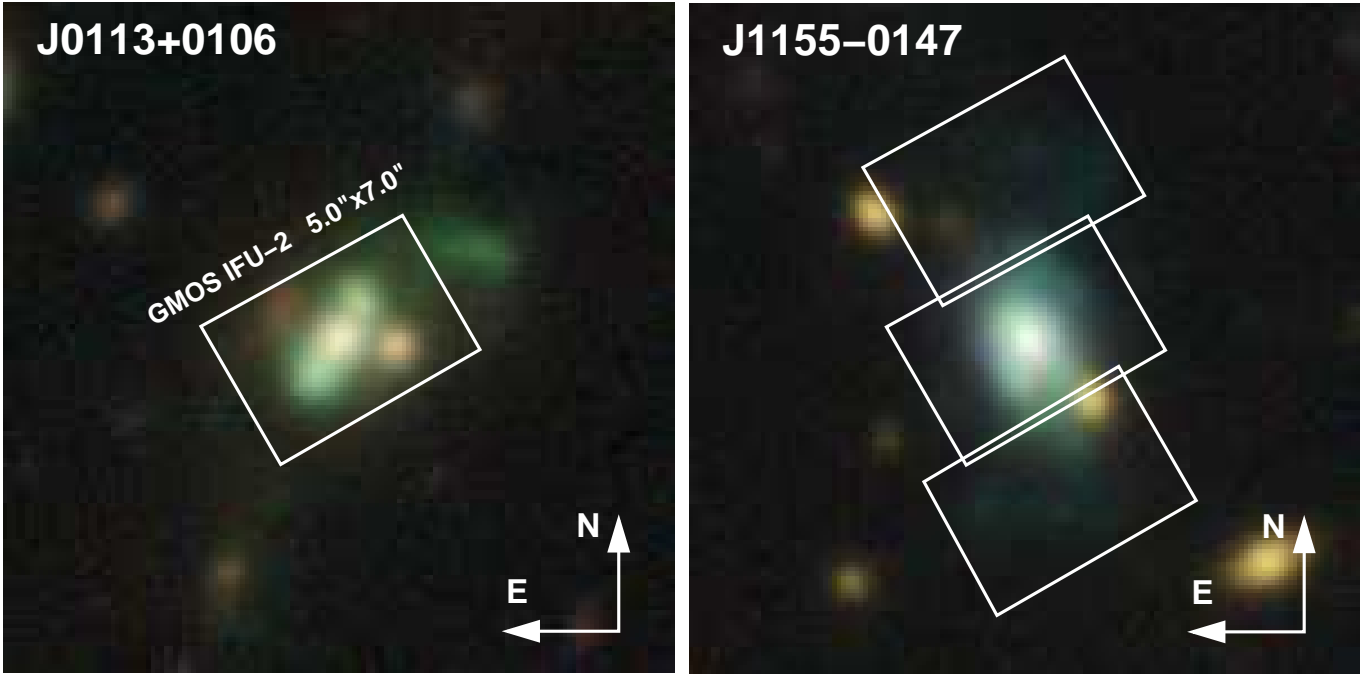}
\caption{\label{ifu_mosaic} 
    Layout of the GMOS 3D spectroscopy observations overlaid on GMOS $gri$ color images, the
    latter measuring 93 and 99 kpc at the source redshifts of J0113 and J1155, respectively.
    Three pointings were used for J1155 to cover the full emission region. 
    The [\ion{O}{III}] emission dominates the $r$-band exposures, rendered green in these images. No offsets are observed between the \chandra detections and the peaks of the optical emission.}
\end{center}
\end{figure*}

\subsection{Gemini/GMOS Observations}
The 3D spectra were obtained using GMOS at the Gemini-South telescope in Chile, using the IFU-2
configuration. The 5\myarcsec0$\times$7\myarcsec0 field of view is sampled with 0\myarcsec2 fibers,
and corresponds to $22\times31$\,kpc at $z=0.3$.

J0113 was observed on 2016 November 06 UT (Gemini program ID GS-2016B-DD-4)
using $4\times1200$\,s integrations in excellent seeing conditions (0\myarcsec3--0\myarcsec4);
the physical resolution corresponds to 1.5\,kpc at $z=0.281$.
The pointing was centered on the nucleus and contains the bulk of the flux including
the two superbubbles (see Figure~\ref{ifu_mosaic}, left panel). The [\ion{O}{III}]$\lambda$5008 line was compromised by a bad detector column that had developed around the time of the observations. The [\ion{O}{III}]$\lambda$5008 line luminosity and principal velocity structures could be reconstructed from the weaker [\ion{O}{III}]$\lambda$4960 line because of the fixed [\ion{O}{III}]$\lambda$5008/4960 line ratio of 2.984 \citep{Sto00} and the good S/N ratio.

J1155 was observed on 2015 March 31 (GS-2014B-Q-63). Due to its large physical extent,
three adjacent pointings were obtained with the IFU-2 (Figure~\ref{ifu_mosaic}, right panel), exposed for $2\times900$\,s each, in $\sim$0\myarcsec7--0\myarcsec8 seeing conditions. The physical resolution is 3.5\,kpc. 

The spectra were processed using Python/IRAF scripts\footnote{Equivalent to the example available from the Gemini data reduction forum:\\{\tt drforum.gemini.edu/topic/gmos-ifu-data-reduction-example-2/}} 
developed for GMOS IFU data. A detailed description of the reduction can be found in \cite{dav15} for SDSS J2240$-$0927, a low redshift LAB in our sample for which we have performed extensive emission line diagnostics. 

The ionized gas kinematics were derived using the [\ion{O}{III}] emission lines. For J1155
both the [\ion{O}{III}]$\lambda$4960 and [\ion{O}{III}]$\lambda$5008 lines were used. For
J0113, the brighter [\ion{O}{III}]$\lambda$5008 line from half the IFU field of view was
projected on a bad detector column and lost, and therefore only the weaker line was used.
For each spaxel, a constant continuum value was subtracted, and then each emission line was
fit with a single Gaussian. If both lines were fit, then we used their fixed line ratio and
required the same velocity and velocity dispersion.

\section{Broadband X-ray Spectral Analysis}\label{sec:broad_ana}

Using {\tt XSPEC}, we simultaneously fit the {\it Chandra} and {\it NuSTAR} spectra, which cover 
the 0.5--10 keV and 3--30 keV bands, respectively. The {\it Chandra} and {\it NuSTAR} spectra are 
binned so that each bin contains at least 15 cts and 100 cts, respectively, driven by the typical 
background rates. Bins are excluded from the spectral analysis if the source signal is below 
$1\sigma$ significance. 
The cross-normalization factor between the {\it Chandra} and {\it NuSTAR} spectra is set to one. 
We confirm that even if the factor is allowed to vary within a typical uncertainty of 10\% \citep{Har13}, 
the fitting results as well as the conclusion do not change significantly.
 
Following previous work by \cite{Kaw16}, who systematically analyzed broadband X-ray spectra of moderately obscured 
($\log(N_{\rm H}$/cm$^{-2})$ = 22--24) AGN, we adopt a base-line model,
\begin{equation}
F(E) = e^{-N_{\rm H}\sigma(E)}P(E) + f_{\rm scat}P(E) + R(E) + G(E)
\end{equation}
in units of photons keV$^{-1}$ cm$^{-2}$ s$^{-1}$. Here, $\sigma(E)$ is the cross-section of photoelectric absorption at energy $E$. Our model uses a cutoff power-law for the absorbed primary X-ray emission, $P(E) \propto E^{-\Gamma}e^{-E/E_{\rm cut}}$, where $\Gamma$ is the photon index. The cutoff energy, $E_{\rm cut}$, is difficult to constrain from the data, and thus fixed to 300 keV, a typical value measured in nearby AGN \citep{Dad08}. The normalization of the primary emission between the \chandra and \nustar spectra are not tied to each other, due to the possible time variability between the two observations. An un-absorbed cutoff power-law model ($f_{\rm scat}P(E)$) is also incorporated as scattered emission, where $f_{\rm scat}$ is the scattered fraction. The last two terms, $R(E)$ and $G(E)$, represent the  reflection continuum from distant, cold matter \citep[``torus"]{Urr95}, and the accompanying fluorescence iron K$\alpha$ line at 6.4 keV, respectively. The torus' reflection component is represented using the {\tt pexrav} model \citep{Mag95}, which calculates reflected emission in an optically thick slab with a solid angle $\Omega$ irradiated by a point source. That is, $R_{\rm ref} \equiv \Omega/2\pi =$ 2 corresponds to the extreme case where the nucleus is covered by the reflector in all directions. 
Initially, the incident flux into the reflector is assumed to be the same as the primary power-law component of the \nustar spectra. The inclination angle to the reflector is fixed at 60$^\circ$. We confirm that even for 30$^\circ$ the best-fit parameters do not change significantly. We assume that the reflected and scattered components did not vary between the \chandra and \nustar observations. The scattering/reflecting matter could be spatially distributed and extend to the pc scale. This suggests that even if the incident emission from the nucleus varies,  the scattered/reflected emission does not change drastically because such matter smears the variability. 
The K$\alpha$ line is described using the delta function ({\tt zgauss} with the line width of 0 eV in {\tt XSPEC}). 
This is because the poor photon statistics and the low instrumental energy resolution do not permit an estimate.
The galactic absorption, $N^{\rm gal}_{\rm H}$, is calculated using the {\tt nh} command in FTOOLS \citep{Kal05}, 
and is applied to the total base-line model.

In the following subsections we describe the details of the spectral analysis. The unfolded spectra and best-fit models are shown in Figure~\ref{fig:src_spec}. The best-fit parameters, observed fluxes, and primary power-law luminosities are listed in Table~\ref{tab:para}. 
To estimate luminosity errors, we make confidence contours between the normalization and photon index of the primary power-law component (Figure~\ref{fig:cont}). Then, we derive the upper and lower boundaries by calculating all the possible luminosities along the contours at the 90\% confidence level. 

\subsection{Notes on J0113}\label{sec:j0113}

The base-line model fits the spectra with $\chi^2/{\rm d.o.f} = 3.3/6$. The AGN in J0113 is heavily obscured with a 
hydrogen column density of $N_{\rm H} = 5.1^{+3.1}_{-3.3}\times10^{23}$ cm$^{-2}$, and we find $\Gamma=2.1^{+0.5}_{-1.2}$. 
The K$\alpha$ line and the scattered emission are insignificant, and their upper limits 
are constrained to be  $A_{\rm K\alpha}< 2.2\times10^{-6}$ photons cm$^{-2}$ s$^{-1}$ ($<1.3\times10^{-14}$ \ergcms) 
and $f_{\rm scat} < 9.9$\,\%, respectively. The reflection strength is estimated to be $R_{\rm ref}=2.7$. 
If we assume that the incident flux into the reflecting and scattering material is 
based on the primary X-ray emission in the \chandra spectrum (instead of \nustar), then $R_{\rm ref} \sim 0.5$ and 
$f_{\rm scat} \lesssim 4$\,\%. Due to the poor statistics, the best-fit parameters in both cases are consistent with each 
other. 

\begin{figure*}[!ht]
\begin{center}\vspace{.8cm}\hspace{.8cm}
\includegraphics[angle=-90,scale=0.33]{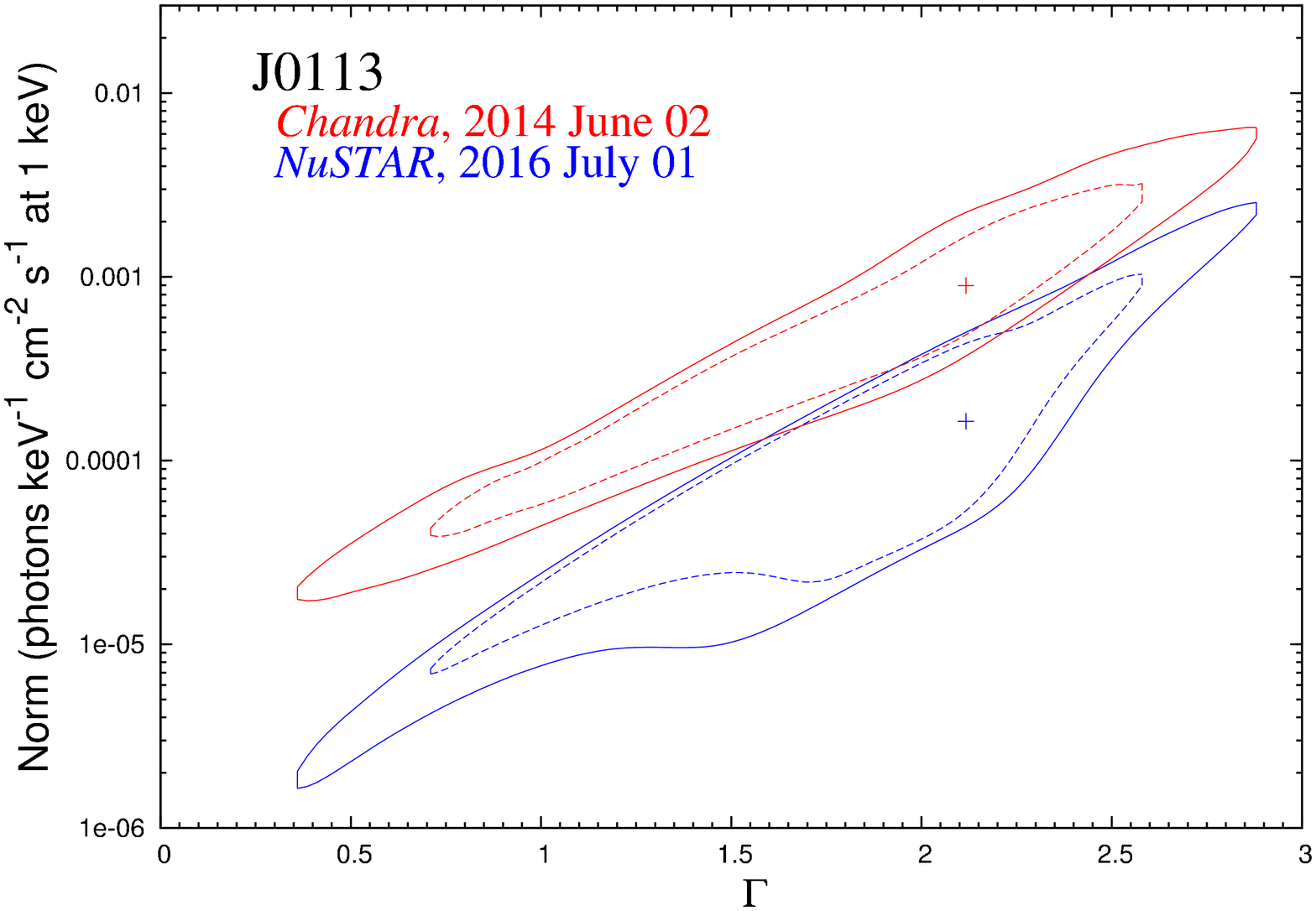}\hspace{.3cm}
\includegraphics[angle=-90,scale=0.33]{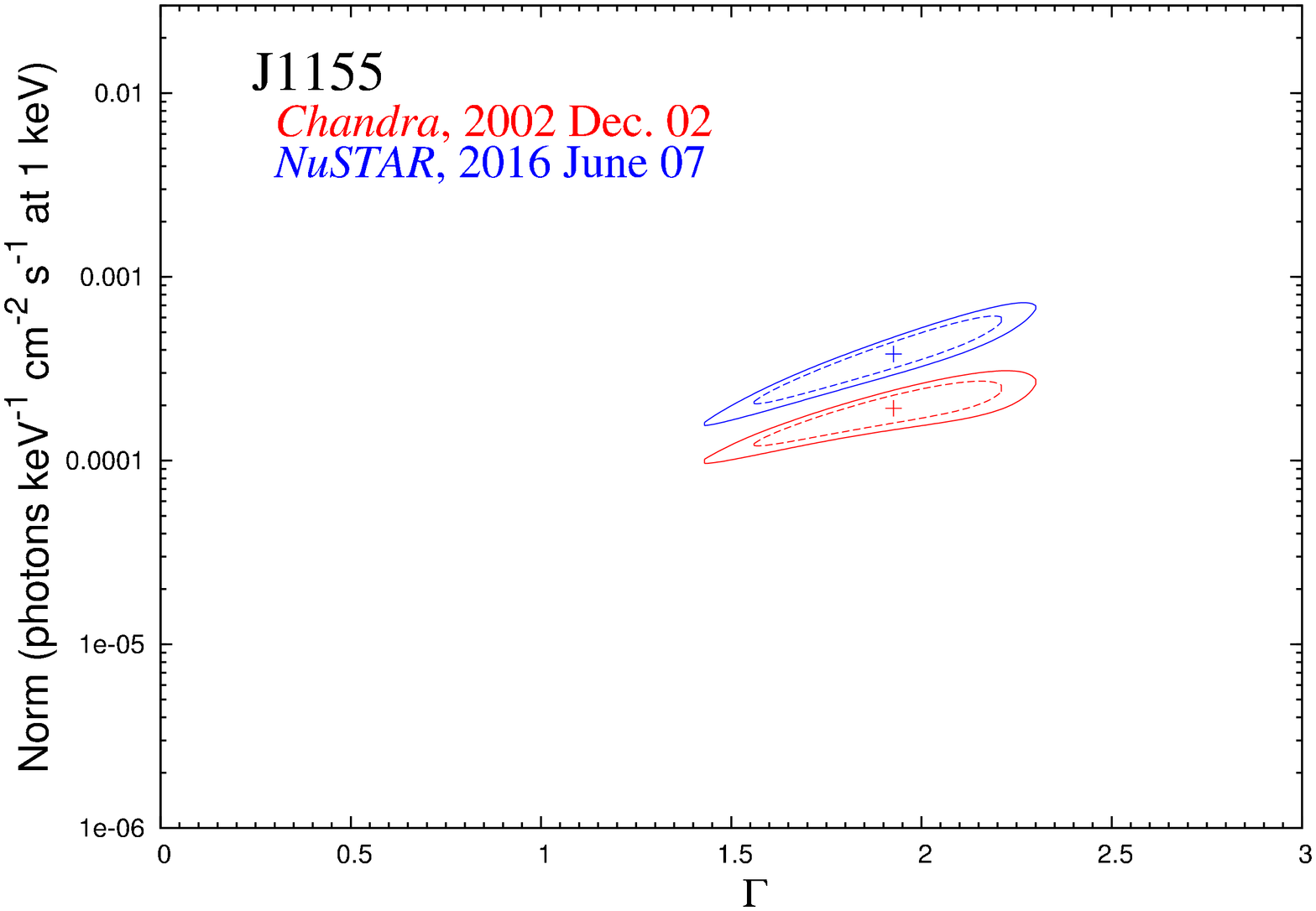}\vspace{-.8cm}
\caption{{\label{fig:cont} 
Confidence contours between the normalization and photon index of the primary power-law component based on 
the simultaneous fit to the spectra from \nustar (blue) and \chandra (red). While each cross denotes the 
best-fit value, the 68\% and 90\% confidence constraints 
are represented with the dashed and solid lines, respectively. 
}}
\end{center}
\end{figure*}

From the confidence contours in Figure~\ref{fig:cont}, the intrinsic luminosities during the {\it Chandra} and {\it NuSTAR} 
observations are $\log (L_{2-10\;{\rm keV}} / {\rm erg\;s}^{-1}) = 44.5^{+0.5}_{-0.6}$ and $43.7\pm0.8$, respectively 
(90\% confidence errors on the luminosities). This AGN has likely faded by one order of magnitude in two years. 

\subsection{Notes on J1155}\label{sec:j1155}

The base-line model reproduces the spectra well with $\chi^2/{\rm d.o.f} = 28.9/35$. We find $\Gamma = 1.9^{+0.3}_{-0.4}$ and $N_{\rm H} = 6.0^{+1.4}_{-1.1}\times10^{22}$ cm$^{-2}$. The K$\alpha$ line is detected at the 90\,\% confidence level with $A_{{\rm K}\alpha} = (1.9\pm1.3)\times10^{-6}$ photons cm$^{-2}$ s$^{-1}$, or $(1.9\pm1.4)\times10^{-14}$ \ergcms.  This supports the existence of a torus. Although the spectra do not require the reflection continuum ({\tt pexrav}) significantly as $R_{\rm ref} = 1.0$ $(< 2.5)$, this continuum emission should be taken into consideration to explain the K$\alpha$ line in a self-consistent way. The scattered emission is significantly detected below 1 keV, and the scattering fraction is $f_{\rm scat} = 1.2^{+1.6}_{-0.9}$\,\%. This falls within a typical range of obscured Seyfert galaxies \citep[0.1--10\,\%; e.g.,][]{Kaw16}. Note that $R_{\rm ref}$ and $f_{\rm scat}$ with respect to the primary emission in the \chandra spectrum are $R_{\rm ref} \approx 2$ and
$f_{\rm scat} \approx 2$\,\%, consistent with those derived above within uncertainties.
Analogously to J0113, we find $\log (L_{2-10\;{\rm keV}}/{\rm erg\;s}^{-1})= 44.0^{+0.1}_{-0.2}$ and $44.3\pm0.1$ for the epochs of the {\it Chandra} and 
{\it NuSTAR} observations, respectively. This AGN has brightened slightly between 2002 and 2016.

\tabletypesize{\scriptsize}
\begin{deluxetable}{lcccccccccc}
\tablecaption{Best-fit Parameters \label{tab:para}}
\tablewidth{0pt}
\startdata \hline 
(1) & Name                                      & J0113              & J1155 \\ \hline 
(2) & $N^{\rm Gal}_{\rm H}$ ($\times 10^{20}$)  & $3.03$             & $2.04$       \\
(3) & $N_{\rm H}$ ($\times 10^{22}$)            & $51^{+31}_{-33}$   & $6.0^{+1.4}_{-1.1}$ \\
(4) & $\Gamma$                                  & $2.1^{+0.5}_{-1.2}$& $1.9^{+0.3}_{-0.4}$ \\
(5) & $A^{Chandra}_{\rm PL}$ $(\times 10^{-4})$ & $9.0^{+29.0}_{-8.8}$  & $1.9^{+0.9}_{-0.8}$ \\
(6) & $A^{NuSTAR}_{\rm PL}$  $(\times 10^{-4})$ & $1.6^{+11.0}_{-1.0}$ & $3.8^{+2.6}_{-1.9}$ \\
(7) & $f_{\rm scat}$                            & $< 9.9$ & $1.2^{+1.6}_{-0.9}$ \\
(8) & $A_{{\rm K}\alpha}\;(\times 10^{-6})$     & $< 2.2$             & $1.9\pm1.3$ \\
(9) & $R_{\rm ref}$                             & $2.7^{\ast}$        & $1.0 (< 2.5)$ \\
(10) & $F^{Chandra}_{\rm 2-10}$ ($\times 10^{-13}$) & $3.6$   & $3.6$ \\
(11) & $F^{NuSTAR}_{\rm 2-10} $ ($\times 10^{-13}$) & $1.3$   & $6.1$ \\
(12) & $\log L^{Chandra}_{\rm 2-10}$ & $44.5^{+0.5}_{-0.6}$& $44.0^{+0.1}_{-0.2}$ \\
(13) & $\log L^{NuSTAR}_{\rm 2-10} $ & $43.7\pm0.8$& $44.3\pm0.1$ \\
(14) & $\chi^2/{\rm d.o.f.}$  & $3.3/6$   & $28.9/35$ \\
(15) & \nustar cts (src/bg) & $212/451$  & $797/717$ \\  
(16) & \chandra cts (src/bg) & $117/1$ & $526/1$ 
\enddata 
\tablecomments{Columns: 
(1) Galaxy name. 
(2) Galactic absorption in units of cm$^{-2}$. 
(3) Intrinsic absorption in units of cm$^{-2}$. 
(4) Photon index of the power-law component.
(5)--(6) Photon fluxes in units of photons keV$^{-1}$ cm$^{-2}$ s$^{-1}$ at 1 keV of 
the primary power-law component in the \chandra and \nustar spectra, 
(7) Scattered fraction in units of \%. 
(8) Total photon flux in the {\tt zgauss} model in units of total photons cm$^{-2}$ s$^{-1}$.
(9) Reflection strength defined as $R_{\rm ref}= 2\pi/\Omega$ in the  {\tt pexrav} model. 
(10)--(11) Observed fluxes in the 2--10 keV band in units of \ergcms. 
(12)--(13) Intrinsic (primary power-law component) luminosities at the galaxy rest frame in the 2--10 keV band. 
(14) Reduced chi-squared over degrees of freedom. 
(15)--(16)  Source and background counts in the \nustar 3--30 keV  and \chandra 0.5--10 keV bands. 
}
\tablenotetext{$\ast$}{The error at the 90\,\% confidence level cannot be constrained within a range, where we allowed the parameter to vary ($R_{\rm ref}=0$--$10$).} 
\end{deluxetable}

\section{Analysis of the optical 3D spectra}\label{sec:Ana3Dspec}
\subsection{J0113: [\ion{O}{III}] and H$\beta$ luminosities}{\label{j0113lum}}
J0113 extends over $7^{\prime\prime}\times13^{\prime\prime}$, featuring two prominent
superbubbles at a projected distance of 5\,kpc on either side of the nucleus.
The bubbles have a diameter of 5--8\,kpc, each. A further outflow is found 24 kpc (North-West) 
from the nucleus and measures 14\,kpc. Judging from its $r$-band brightness and the
[\ion{O}{III}] equivalent line width of the central region, this outflow contributes
$15$--$20$\% to the total [\ion{O}{III}] luminosity, which we disregard in the further
analysis. For more details see Figure~\ref{ifu_mosaic} in this paper, and Appendix
A4 in \cite{Sch16}.

The total [\ion{O}{III}]$\lambda$5008 flux in the IFU-2 data is
$(4.36\pm0.44)\times10^{-14}$ \ergcms, equivalent to an observed (reddened) luminosity of
$1.1\times10^{43}$ \ergs. The archival spectrum from the Sloan Digital Sky Survey (SDSS)
has a mean global H$\alpha$/H$\beta$ line ratio (Balmer decrement) of 4.4, indicative
of significant dust extinction. Using a standard $R=3.1$ dust extinction curve
\citep{car89,osf06}, an intrinsic H$\alpha$/H$\beta$ line ratio of 3.0, we find
${\rm log}\,(L^c_{[\ion{O}{III}]} / {\rm erg\;s}^{-1}) = 43.6\pm0.1$ for the dereddened
luminosity. Likewise, we measure a corrected H$\beta$ line luminosity of
${\rm log}\,(L^c_{{\rm H}\beta} / {\rm erg\;s}^{-1}) = 42.9\pm0.1$. The 1$\sigma$ errors
include the uncertainty of the flux calibration and allow for a relative uncertainty of
10\% in the intrinsic H$\alpha$/H$\beta$ line ratio; the latter is subject to metallicity
and hardness of the ionizing spectrum \citep{gas84}. Possible variations of the internal
dust extinction remained unaccounted for because we do not have resolved H$\alpha$ maps
at hand \citep[for examples see][]{Sch13,dav15}.

\begin{figure*}[ht]
  \begin{adjustbox}{addcode={\begin{minipage}{\width}}{\caption{\label{velocitymaps} 
    Gas velocity (left), velocity dispersion (center) and line flux (right) for J0113
    (top) and J1155 (bottom).
    The line flux is also overplotted by the black contours
    in the velocity and dispersion panels. J0113 is dominated by a bipolar outflow,
    and a central cloud of rotating gas. The outflow axis and the rotation axis are
    aligned. J1155 has a very concentrated line flux and drives a narrow outflow over
    20 kpc, with constant radial velocity. The masking corresponds to approximately a
    S/N cut of 5. Our IFU data are comparatively shallow; the gas in J0113 extends
    well beyond the plot frame, and the gas in J1155 fills the frame.
      }\end{minipage}},rotate=90,center}
      \includegraphics[scale=0.9]{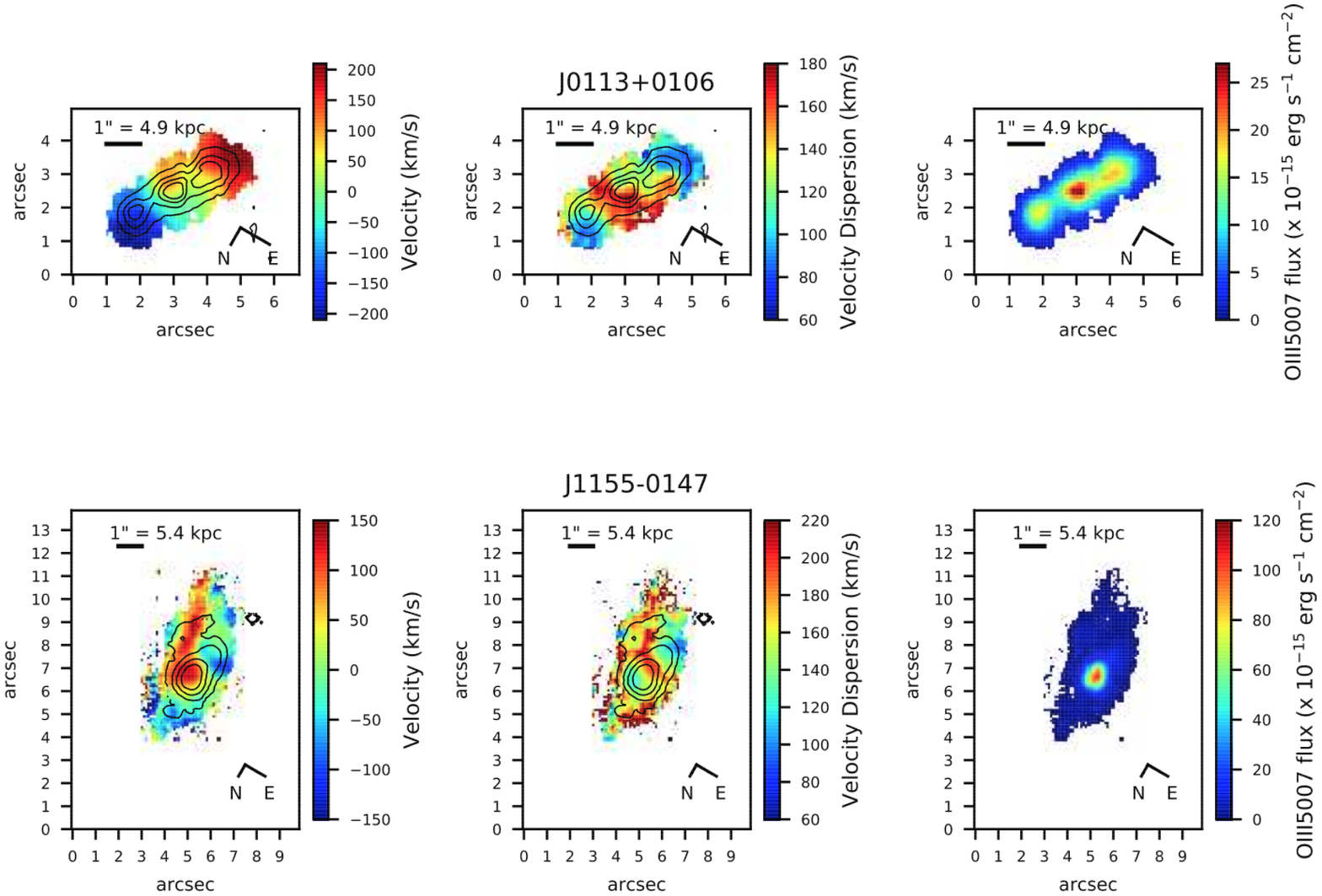}%
  \end{adjustbox}
\end{figure*}

\subsection{J0113: Gas kinematics}
The kinematic maps in Figure~\ref{velocitymaps} reveal a bipolar outflow.
Relative radial velocities increase linearly from the nucleus
to about 120 km\,s$^{-1}$ at the position of both superbubbles at 5 kpc distance.
The Northern superbubble is blue-shifted, and the Southern bubble redshifted. 
Velocities then continue to increase in the same fashion beyond the bubbles,
exceeding 200 km\,s$^{-1}$ at 10 kpc from the nucleus, where the S/N becomes too
low for our analysis. The outflow does not appear to encounter significant
mechanical resistance from the interstellar medium, and its motion is not yet
governed by gravitation. The measured radial velocity is fairly low, suggesting
that most of the motion happens perpendicular to our line-of-sight.
 
Another feature is that the gas South-West (North-East) of the nucleus
is redshifted (blueshifted). This velocity gradient is perpendicular to the
outflow and with a lower amplitude of about 70 km\,s$^{-1}$. There is no indication
of an outflow in this direction, and we interpret this as rotational motion of
the central cloud of ionized gas.

The velocity dispersion is low (Figure \ref{velocitymaps}, upper middle), and it is
remarkable that the nucleus is not detectable in this map. We note a band of increased
dispersion, $150$--$180$ km\,s$^{-1}$, that appears to gird the nucleus and the bipolar
outflow near its equatorial plane. In the two superbubbles and beyond,
the dispersion drops to $\lesssim100$ km\,s$^{-1}$, lower than in other
outflow systems. We discuss these properties
in Sections \ref{discussion-kinematics} and \ref{discussion-dispersion}.

\subsection{J1155: [\ion{O}{III}] and H$\beta$ luminosities}{\label{j1155lum}}
J1155 has the highest X-ray flux in our sample. The optical nebula in the
$gri$ broadband images of \cite{Sch16} extend over
$9^{\prime\prime}\times19^{\prime\prime}$ or more than 60\,kpc, and fragments into
numerous smaller clouds of 3--7\,kpc diameter scattered around a very bright
center measuring 5\,kpc in diameter. For more details see
Figure~\ref{ifu_mosaic} in this paper, and Appendix A6 in \cite{Sch16}.

We have no spectrum yet of J1155 that includes the H$\alpha$ line, and
therefore we cannot determine a dereddened [\ion{O}{III}] luminosity. The observed
total line flux including calibration errors is $(8.0\pm0.8)\times10^{-14}$ \ergcms,
corresponding to ${\rm log}\,(L_{[\ion{O}{III}]} / {\rm erg\;s}^{-1}) = 43.4\pm0.05$.
If the Balmer decrement was a moderate 3.5, then
${\rm log}\,(L^c_{[\ion{O}{III}]} / {\rm erg\;s}^{-1}) = 43.6\pm0.1$ and
${\rm log}\,(L^c_{{\rm H}\beta} / {\rm erg\;s}^{-1}) = 42.5\pm0.1$.
If the Balmer decrement was as high as for J0113, then
${\rm log}\,(L^c_{[\ion{O}{III}]} / {\rm erg\;s}^{-1}) = 43.9\pm0.1$.
For the rest of this paper we adopt the more conservative, lower luminosities
associated with less extinction.

\subsection{J1155: Gas kinematics}
J1155 has a very complex kinematic structure and [\ion{O}{III}] morphology. The
line emission is strongly concentrated in the center of the galaxy, and with low
velocity dispersion. Ionized gas extends out to $\sim25$ kpc in these shallow IFU
data, creating a patchy appearance and complex velocity and velocity dispersion
maps. Particularly interesting is a long stream of gas connected to one side of
the nucleus, stretching over 20 kpc and receding with a constant radial velocity
of $\sim120$ km\,s$^{-1}$. The width of the stream is barely resolved in the
data (Figure \ref{velocitymaps}, lower left); the velocity dispersion is increased
further away from the nucleus. The remainder of the surrounding gas appears to be
passively illuminated by the AGN.

Like J0113, the velocity dispersion in J1155 is fairly low, between
$100$--$200$ km\,s$^{-1}$ across the entire analysis area of $30\times15$ kpc.
The nucleus is also not detectable (Figure \ref{velocitymaps}, lower middle).

\section{Discussion}\label{sec:results}

\subsection{X-ray properties}

\subsubsection{Constraint on X-ray luminosity}

\chandra is sensitive to soft energy X-ray photons in the $\sim 0.7$--$9$ keV restframe for 
objects at $z\sim0.3$. Hence, it is difficult to put a strong constraint on the intrinsic 
luminosity when the source is highly  obscured. To 
show this fact, we derive luminosity errors in the case of that only the \chandra spectrum 
is available. As an example, we consider the J0113 case. We apply the same method as described in 
Section~\ref{sec:broad_ana} to the error calculation except for allowing the absorbing column 
density to change up to $\log N_{\rm H} = 24$ in the fitting to avoid extreme case.
The resultant lower and upper boundaries 
of the primary power-law luminosity become $-0.8$ dex and $+1.5$ dex.  
These uncertainties are larger than obtained with the simultaneous fitting to 
the \nustar and \chandra spectra 
($-$0.6 dex and $+$0.5 dex) possibly because the absorption column density is not well determined. 
This shows that the additional \nustar spectrum plays an important role to constrain 
the luminosity.

\subsubsection{Narrow iron-K$\alpha$ line}\label{sec:iron_line}

The narrow K$\alpha$ line at $\approx$ 6.4 keV may originate from the torus around the nucleus, and was detected 
significantly only for J1155. Its equivalent width with respect to the reflection continuum is EW$=760^{+830}_{-580}$ eV 
(EW$< 7900$ eV for J0113). Here, only an uncertainty of the iron line flux is incorporated into the error calculation 
because the lower limit of the reflection strength is not constrained. The EWs are consistent with a numerical calculation 
of the reflected emission for optically thick, cold matter \citep[1--2 keV;][]{Mat91}. On the other hand, the EWs of J0113 
against the total continuum observed with \chandra and \nustar are $<240$ eV and $<740$ eV, whereas those of J1155 are 210 
eV and 130 eV, respectively. All observed EWs are consistent with those found in typical obscured AGN 
\citep[e.g.,][]{Bri11a,Kaw16}. We conclude that our low-$z$ LABs and typical Seyfert galaxies may have the same origin of 
the K$\alpha$ line.



\subsubsection{Scattering fraction}

We observe significant scattered emission ($f_{\rm scat} \sim 1.2$\%) for J1155,
which implies that J1155 is not heavily buried in the dust or gas obscuration.
In comparison, heavily buried AGN (as suggested from their strong reflected emission)
show rather low scattering fractions of $f_{\rm scat}<0.5\,\%$ \citep[e.g.][]{Ued07,Egu09}.
We conclude that the emission from the nucleus in J1155 is relatively unobscured,
and that the AGN contributes over a large solid angle to the ionization of the LAB.

By contrast, we cannot find significant scattered emission in J0113 ($f_{\rm scat}<9.9$\,\%), 
suggesting the presence of a buried AGN and also a smaller contribution to the ionization. 

\subsection{Properties from the optical 3D spectra}\label{3dspecdiscussion}
\subsubsection{Comparison samples}
How are the galaxy-scale outflows in this study different from other systems with similar properties?
Our targets were selected from SDSS-DR8 \citep{Aih2011}
solely based on their broad-band photometry (excess of $r$-band flux due to
redshifted, very high [\ion{O}{III}] emission at $z\sim0.3$). We also applied
a lower threshold on their radius to select objects with extended outflows.
Both our targets are radio-quiet and ionized by type-2 AGN \citep{Sch13}.

We compare with the 16 radio-quiet type-2 AGN of \cite{Har14}, selected from SDSS
because of their high spectroscopic [\ion{O}{III}]$\lambda$5008 flux. \cite{Har14}
also required the presence of a broad [\ion{O}{III}] component. Typical redshifts
are $z\sim0.1$. Our other comparison sample is that of \cite{Liu13}, who selected
11 type-2 AGN with galaxy-scale outflows (at $z\sim0.5$) in SDSS based on high
[\ion{O}{III}] fluxes. \cite{Liu13} complement their sample with 3 radio-loud
AGN to study the impact of jets on the outflow systems. Like us, \cite{Har14}
and \cite{Liu13} used Gemini/GMOS to obtain the 3D spectra, in comparable seeing
conditions. The depth of the spectra are approximately similar to ours of J0113.
Our data of J1155 is much shallower though, by about a factor $\sim10$ in effective
exposure time (short integrations; full moon, compared to dark sky conditions
for the comparison data).

\subsubsection{Ionized gas mass}
We derived H$\beta$ logarithmic line luminosities of 42.9 and 42.5 for J0113 and J1155,
respectively (Sections \ref{j0113lum} and \ref{j1155lum}). Assuming the gas is purely
photoionized, the gas mass $M$ associated with the H$\beta$ emission is estimated as

\begin{equation}
  M=\frac{L_{{\rm H}\beta}}{4\pi j_{{\rm H}\beta}}\,n_{\rm e}\,m_{\rm p}\,,
\end{equation}
where $4\pi j_{{\rm H}\beta}$ is the emission per unit volume, $n_{\rm e}$ the electron
number density and $m_{\rm p}$ the proton mass. Taking $4\pi\,j_{{\rm H}\beta}$ from
\cite{osf06} \citep[their table 4.4; see also][]{Sto15} for $T=10^4$\,K and
$n_{\rm e}=100$ cm$^{-3}$, we rewrite

\begin{equation}
  M = 6.77\times10^8\,M_\odot\,
  \frac{L_{{\rm H}\beta}}{10^{43}\,{\rm erg\,s^{-1}}}\,
  \left( \frac{n_{\rm e}} {100\,{\rm cm^{-3}}}\right)^{-1}\;.
  \label{hbmass}
\end{equation}

Individual measurements of $n_{\rm e}$, derived from the
[\ion{S}{II}]$\lambda$6719,33 line ratio, yield several 100 cm$^{-3}$ in the inner
regions of narrow-line regions \citep[see e.g.][]{Nes06,dav15}. Using line ratios
which are sensitive at higher densities, $n_{\rm e}\sim10^{3}$ ($10^{4-5}$) cm$^{-3}$
were found for the narrow (broad) lines of another luminous quasar \citep{Hol11}.
In the outskirts, densities drop below 100 cm$^{-3}$, where the outflows have thinned
and are difficult to detect (and where line ratio methods break down because the
collision effects diminish). Our spectrum of J0113 does not include the
[\ion{S}{II}] doublet; however, there is an SDSS spectrum taken with a
3$^{\prime\prime}$ fiber centered on the nucleus,
including the flux from the nucleus and most of the superbubbles. Fitting the [\ion{S}{II}]
doublet with a simple double Gaussian with identical line widths, we derive a line ratio of
1.24, i.e., a mean $n_{\rm e}\sim180$ cm$^{-3}$. We expect the actual electron densities
to deviate significantly from this number, because outflows are known to be clumpy
\citep{Nes06,Gre11,dav15}.

Using eq. (\ref{hbmass}), we find $5.4\times10^8\,M_\odot$ and $2.0\times10^8\,M_\odot$
for J0113 and J1155, respectively, for a fixed value of $n_{\rm e}=100$ cm$^{-3}$. This
is near the upper end of the gas masses of $(0.2$--$4)\times10^8\,M_\odot$ found by
\cite{Har14}, using the same density normalization. The outflows of \cite{Liu13}
entrain higher masses, $(2$--$10)\times10^8\,M_\odot$ (again for $n_{\rm e}=100$ cm$^{-3}$).

  We note that \cite{Har14} and \cite{Liu13} have a numeric prefactor that is erroneously
  about a factor 4 too high compared to ours in eq. (\ref{hbmass}). This is caused by an
  incorrect calculation of the recombination coefficient $\alpha^{\rm eff}_{{\rm H}\alpha}$
  in eq. (1) of \cite{Nes06} (N. Nesvadba, private communication). Taking this into account, 
  the ionized gas masses of J0113 and J1155 are among the highest measured in type-2 AGN at
  $z\lesssim0.6$.

\subsubsection{\label{discussion-kinematics}Gas velocities}

Assuming a typical, true outflow velocity of 1000 km\,s$^{-1}$ for the [\ion{O}{III}] gas
\citep{Har14}, it would take the two superbubbles in J0113 at least $\sim10^7$ years to
reach their observed positions. However, a sustained high AGN output over such a long time 
is implausible, as single AGN duty cycles are expected to be comparatively short 
\citep[$10^{4-6}$ yr; e.g.,][]{Sch10,Sch15,Sun17} and followed by longer low states.

\begin{figure}
\includegraphics[width=1.0\hsize]{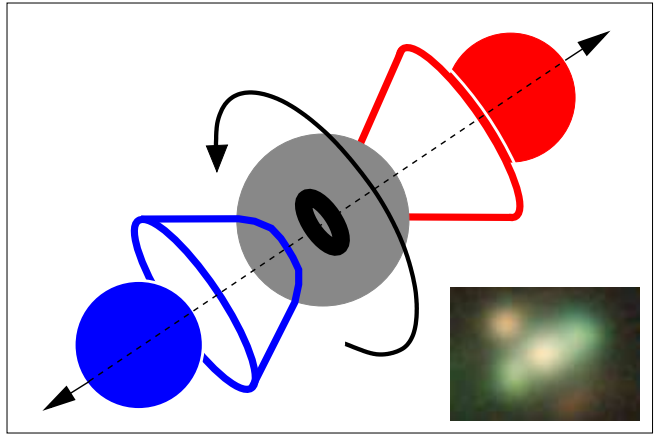}
\caption{\label{toymodel} 
    Model of J0113, showing the system's orientation, outflow geometry and main kinematics
    in our IFU data. The blue and red spheres illustrate the approaching and receding
    superbubbles, respectively. The gray sphere represents the central ionized cloud of gas
    that rotates in the direction indicated. The rotation axis coincides with the direction
    of the outflow. The AGN is obscured from our view by a dusty torus. The two half cones
    indicate how the system lies in the plane of the sky (they are not meant to represent
    ionization cones). The inset shows the optical GMOS image in $gri$ filters.
}
\end{figure}

What happens with such an outflow during the low states of the AGN? \cite{Kin11}
calculate the internal properties of outflows and their dynamics. They find that the
thermal energy contained in an energy-driven outflow can support its expansion for
periods 10 times longer than the actual quasar phase that launched the outflow.
However, their numerical results do not show that the velocity in such a stream 
is increasing over kpc-scales from the nucleus, as we observe in J0113. In a more recent
work, \cite{Zub16} study the evolution of outflows taking into account recurrent AGN
duty cycles lasting $5\times10^4$ years. In these calculations, the recurrent duty
cycles manage to accelerate existing outflows after about $\sim10^6$ years, and out to
radii larger than 10 kpc. \cite{Zub16} also find that the AGN should switch off once
sufficient energy has been injected to unbind all gas; this is not (yet) the case in
J0113 as our X-ray observations still reveal a very powerful AGN (one that has albeit likely 
faded in the last two years). Other processes could also explain the
apparent acceleration. J0113 is weakly detected in the radio VLA FIRST survey 
\citep[1.18 mJy;][]{Sch13}, suggesting that a former radio jet could have played a role 
in shaping the current velocity profile. However, the survey's spatial resolution and
depth \citep{Whi97} are insufficient for further conclusions.
Alternatively, differential velocities in the stream could be at
play (faster components reach further in the same amount of time). However, \cite{Liu13}
argue that in this case the velocity dispersion should increase as well with increasing
distance, yet we observe a decrease in dispersion. We conclude that with the current
data at hand, recurrent outflows are the best explanation for the peculiar velocity
profile in the outflow of J0113.

\cite{Fau12} have shown that an outflow preferentially breaks out from a denser disk or
sphere along paths of least resistance. The hot buoyant gas would then inflate the
bubbles, and bipolar systems should be produced eventually by many powerful AGN.
J0113 seems prototypical in this respect; we condense our observational results in a
simplistic graphical representation in Figure \ref{toymodel}.

J1155, on the other hand, could not be more different. While we do trace a 20 kpc long
linear feature of enhanced recession velocity (Figure \ref{velocitymaps}, lower left),
it is not particularly present in the line flux map (Figure \ref{velocitymaps}, lower
right). The [\ion{O}{III}] nebula fragments into numerous blobs at all azimuthal angles
around the AGN, which do not appear to be connected to any currently active outflow, nor
to any of the member galaxies of the small group that hosts J1155 in its center (see also
Section~\ref{discussion-dispersion} and Figure~\ref{ifu_mosaic}). We think that J1155 represents
a more advanced stage, where the
ejected materials from various duty cycles have accumulated in the halo, but have not
yet fully dispersed. Currently, J1155 experiences another powerful accretion event,
as indicated by the very high central [{O}{III}] surface brightness and its high X-ray
luminosity.

\begin{figure}
\includegraphics[width=1.0\hsize]{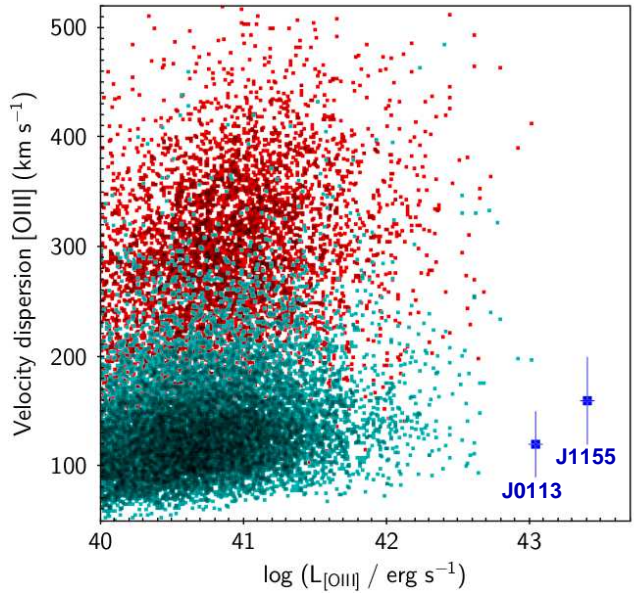}
\caption{\label{mullaney_oiii_fwhm}
  [\ion{O}{III}] velocity dispersion versus (reddened) luminosity for the type-2 AGN
  sample of \cite{Mul13}. The bright blue and red data clouds represent the narrow and 
  broad components, respectively, of the line fits by \cite{Mul13}. Overplotted are our two targets,
  where the vertical error bars represent the actual scattering over the IFU field of
  view. J0113 and J1155 have higher observed [\ion{O}{III}] luminosities than any other
  type-2 AGN in the comparison sample, and broad line components are absent in them.
  Note that \cite{Mul13} originally reported the lines' FWHM; for comparison with our
  data we converted them to velocity dispersions assuming Gaussian line profiles.}
\end{figure}

\subsubsection{\label{discussion-dispersion}Velocity dispersion}
What comes to attention with our IFU data are the fairly low velocity dispersions
in both systems, in particular for their nuclei (see Figure \ref{velocitymaps}). Near the
center of J0113 we observe $150$--$180$ km\,s$^{-1}$, and for the two superbubbles and
beyond the dispersion drops to $100$ km\,s$^{-1}$ and below. For J1155, we observe a mean
dispersion of 160 km\,s$^{-1}$. In Figure~\ref{mullaney_oiii_fwhm} we compare our targets
against the type-2 sample of \cite{Mul13}. We recall that our data of J1155 are shallow,
and therefore any broad components in the outskirts of this nebula would remain
undetected; consequently, we would underestimate its line FWHM.

The low velocity dispersions show that the gas is little turbulent, in particular for
J0113. There, it is flowing away from the AGN in a radial fashion, probably still
pushed by recurrent duty cycles and its internal thermal energy. If powerful earlier
outflows have ``cleared'' the escape path already, then the current stream would find
less resistance and thus be less turbulent. Indeed, our imaging data show the presence
of older outflows that must have traveled along the same paths.

\cite{Liu13} find high ($\sim500$--$1500$ km\,s$^{-1}$) velocity dispersions in
the inner $5$--$10$ kpc of their sample, which in about half of the cases drop
to $100$--$200$ km\,s$^{-1}$ at $10$--$15$ kpc distance. Like us, they argue
that recurrent outflows could have cleared the escape path. Additionally, \cite{Liu13}
explain that the narrow opening angle of the outflow (once it managed to break out)
would leave less space for projected velocities, naturally decreasing
the observed velocity dispersions. A noticeable difference between their and
our data is that J0113 and J1155 have low velocity dispersions \textit{everywhere},
even near the nucleus (our data of J0113 has a physical resolution of 1.5 kpc).
Perhaps the more turbulent central components are simply hidden by dust.
The enhanced Balmer decrement of 4.4 (integrated over 15 kpc by the
SDSS fiber) shows that at least some part of the ionized gas must be severely
affected by absorption, which goes in line with the high column density of
$N_{\rm H}/{\rm cm}^2) = 5.1\times10^{23}$ cm$^{-2}$(J0113).

J1155 also reveals a rather low velocity dispersion, about 160 km s$^{-1}$
averaged over our IFU data. Hardly any of its numerous extended features
appear to be connected to actively driven outflows. This material was
probably ejected during several AGN bursts in the past, and is now passively
ionized by the AGN. A notable environmental difference with respect to
J0113 is that J1155 is not isolated, but located at the center of a small
galaxy group with $M_{200}\sim1.3\times10^{13}\,M_\odot$. \cite{Sch16} speculated
that the unusual morphology of this system could be due to interaction with
infalling gas from an intragroup medium (undetected in X-rays). Alternatively,
the surrounding inner material could have a lower filling factor, offering more
escape paths for any buoyant hot gas. That would also be consistent with the
lower absorbing X-ray column density.

\subsection{Comparison with the high-$z$ LAB `B1'}
One of the best-studied (and most luminous) LABs is SDSS J2143$-$4423 (also known as `B1',
$z=2.38$). Its Ly$\alpha$ emission extends over $30\times100$ kpc \citep{Fra01},
and \cite{Pal04} report a Ly$\alpha$ line flux of $1.35\times10^{-15}$ \ergcms, or 
log$(L_{{\rm Ly}\alpha}/{\rm erg\;s}^{-1})=43.8$. \cite{Ove13} measure a dereddened
[\ion{O}{III}] luminosity of log$(L_{\rm [\ion{O}{III}]}/ {\rm erg\;s}^{-1})=43.4$--$44.1$,
subject to uncertain extinction correction. The optical nebula extends over
$32\times40$ kpc. The [\ion{O}{III}] line FWHM 
within 10 kpc of the nucleus is $600$--$800$ km s$^{-1}$, and drops to 200 km s$^{-1}$ 
beyond 10 kpc \citep[similar to many of the targets studied by][]{Liu13}. 
The radial gas velocities in B1 are low, mostly within $\pm50$ km s$^{-1}$, and the
[\ion{O}{III}] line flux is strongly concentrated. Any substantial bulk 
motions must happen perpendicular to the line of sight. \cite{Ove13} argue that B1
is powered by an AGN that is undetected in X-rays.

Apart from the high velocity dispersion, B1 seems quite similar to J1155.
Both show a high concentration of the line flux, surrounded by a very extended
($\gtrsim 20$ kpc) disk of much lower surface brightness with somewhat patchy
structure. The [\ion{O}{III}] luminosities are comparable, and J1155 is about twice as 
large (judging from our deep imaging data). Its Ly$\alpha$ luminosity estimated from {\it GALEX} 
far-UV data \citep[][]{Sch16} is the same as that of B1.
The diameter of the Ly$\alpha$ emission of J1155, and the accurate Ly$\alpha$
luminosity, are subject of our currently active {\it HST} proposal. J0113 has similar
luminosity and extent as J1155, but its estimated Ly$\alpha$ luminosity is
three times lower than that of B1.

We conclude that the recently discovered low-$z$ LABs match high-$z$ LABs in terms
of luminosity and physical extent. Like B1, our targets host powerful AGN.
At this point it is too early to speculate about the morphological similarities
between B1 and J1155. While the formation of J0113 with its characteristic
bipolar outflow and other signs of historic accretion events is quite
clear cut \citep[see][]{Fau12,Zub16}, the formation and current state of J1155
remains unclear and requires further observations (in particular optical line
diagnostics and deeper data).

\subsection{[\ion{O}{III}] morphologies and AGN flickering}
There is increasing evidence for AGN flickering (duty cycles lasting $10^{4-5}$ years,
amplitudes of orders of magnitude), obtained by the identification of ionization
echoes \citep[e.g.][]{Sch10,Sch16,Ich16,Kee17}, by means of statistical arguments
and outflow rates \citep[e.g.][]{Sch15,Sun17}, the transverse proximity effect
in the Ly$\alpha$ forest \citep[][albeit on longer scales of $10^6$ years]{Kir08},
and also on the theoretical side \citep[e.g.][]{Hop10,Nov11,Sij15,Zub16}.

The typical time scales of AGN flickering are on the same order as the light
crossing times of galaxy-scale outflows extending over 10 kpc and more. The
optical appearance of these extended narrow-line regions is invariable over a
human life time given their kinematic time-scale of Myrs, and the limited
physical resolution of current instrumentation. However, because of their large
extent, these nebulae reflect the recent history of the AGN, and this affects
how we interpret their observed morphologies.

First of all, the effective recombination time scale for the O$^{++}$ ion is very
short. Using the calculations of \citep{Bin87},
\begin{equation}
t_{\rm rec}({\rm O}^{++}) = 158\,{\rm yr}\,\left(\frac{n_{\rm e}}{100\,{\rm cm}^{-3}}\right)^{-1}\,.
\end{equation}
Accordingly, the response of the [\ion{O}{III}] line to a change in ionizing radiation
can be considered instantaneous for observations with a physical resolution of
$\sim 1$ kpc. An outward propagating wave of ionizing radiation (or a shut-down
thereof) would alter the nebulae's surface brightness.

Secondly, we know from the \textit{Changing Look} quasars \citep{Lam15,Mac16,Run16} that AGN 
may change their line-of-sight obscurations on scales of years, affecting an ionization
cone's opening angle (e.g. due to intervening dust, structural changes in the torus in response
to a change of bolometric luminosity). Spin precession of the SMBH accretion disk happens
on scales of $10^{3-7}$ years \citep{LuZ05} and may cause illumination (ionization) patterns
on a screen of gas extending over 10s of kpc.

Lastly, absence of line emission in some area does not imply absence of gas.
These areas could be shielded from ionization by intervening dust. In turn,
absence of line emission does not mean that there is no ionizing radiation;
and AGN may well shine into empty volumes.

It is tempting to link the optical morphologies from our IFU data with the structural
parameters of the central AGN derived from our X-ray analysis. Indeed, the consistencies
are remarkable. From the X-ray data of J0113 we infer an AGN that is buried in a torus
with a small opening angle. Accordingly, this AGN is able to photoionize gas within
a relatively narrow ionization cone, only, matching the collimated bipolar outflow.
However, the highest surface brightness of the line emission is located in the central
$2$--$4$ kpc (upper right in Figure \ref{velocitymaps}), showing that ionizing radiation
can leak into other directions, too. In J1155, we find a weakly obscured AGN capable
of ionizing gas over a wide solid angle, consistent with the optical nebula ionized
at all azimuthal angles. However, our sample is highly limited (two), and as we cautioned
above, the observable nebular morphologies are subject to outflow histories, and 
variabilities in obscuration, luminosity, and illumination direction. 

\begin{figure*}[!ht]
\begin{center}
\includegraphics[scale=0.595]{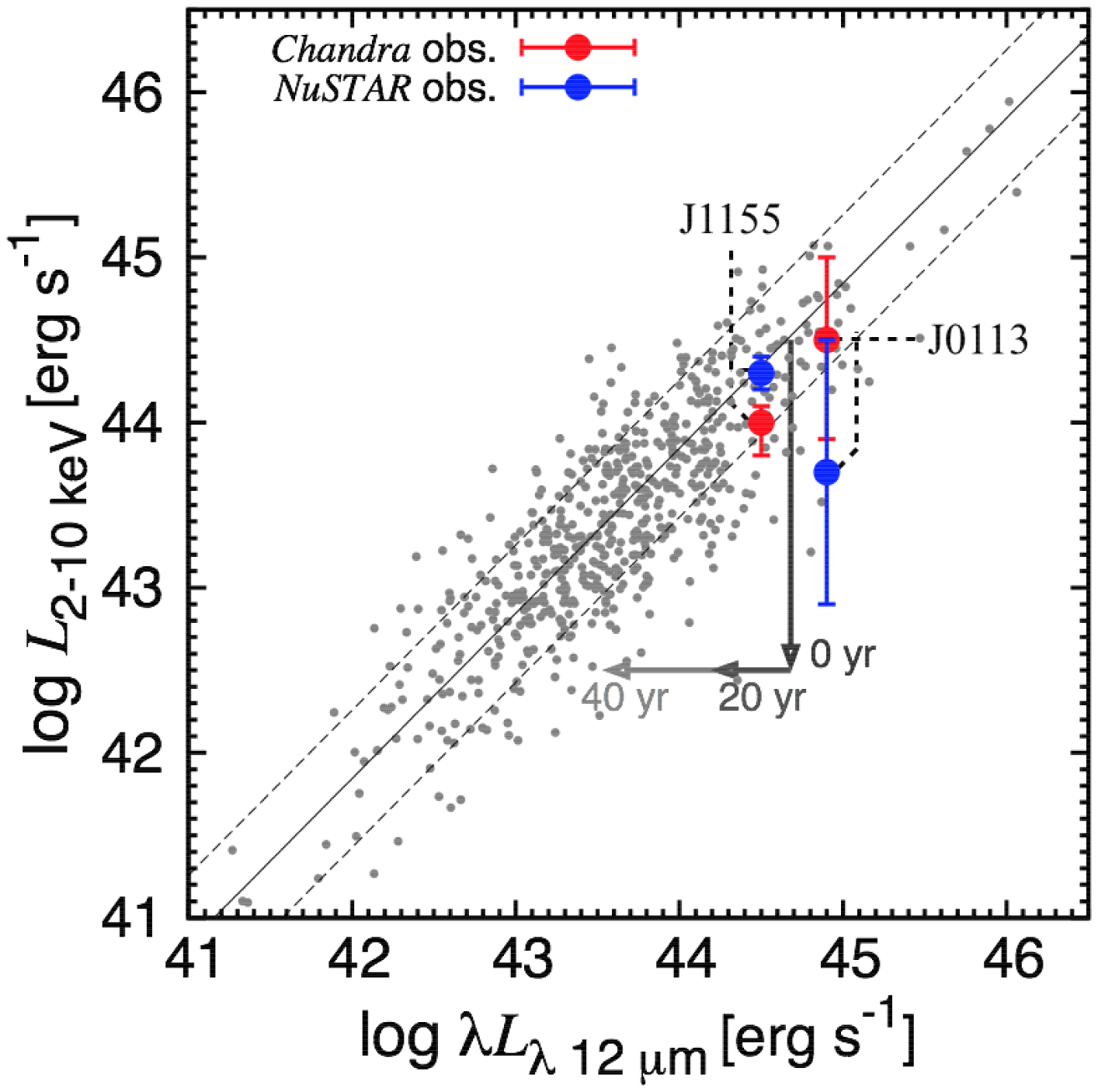}
\includegraphics[scale=0.4]{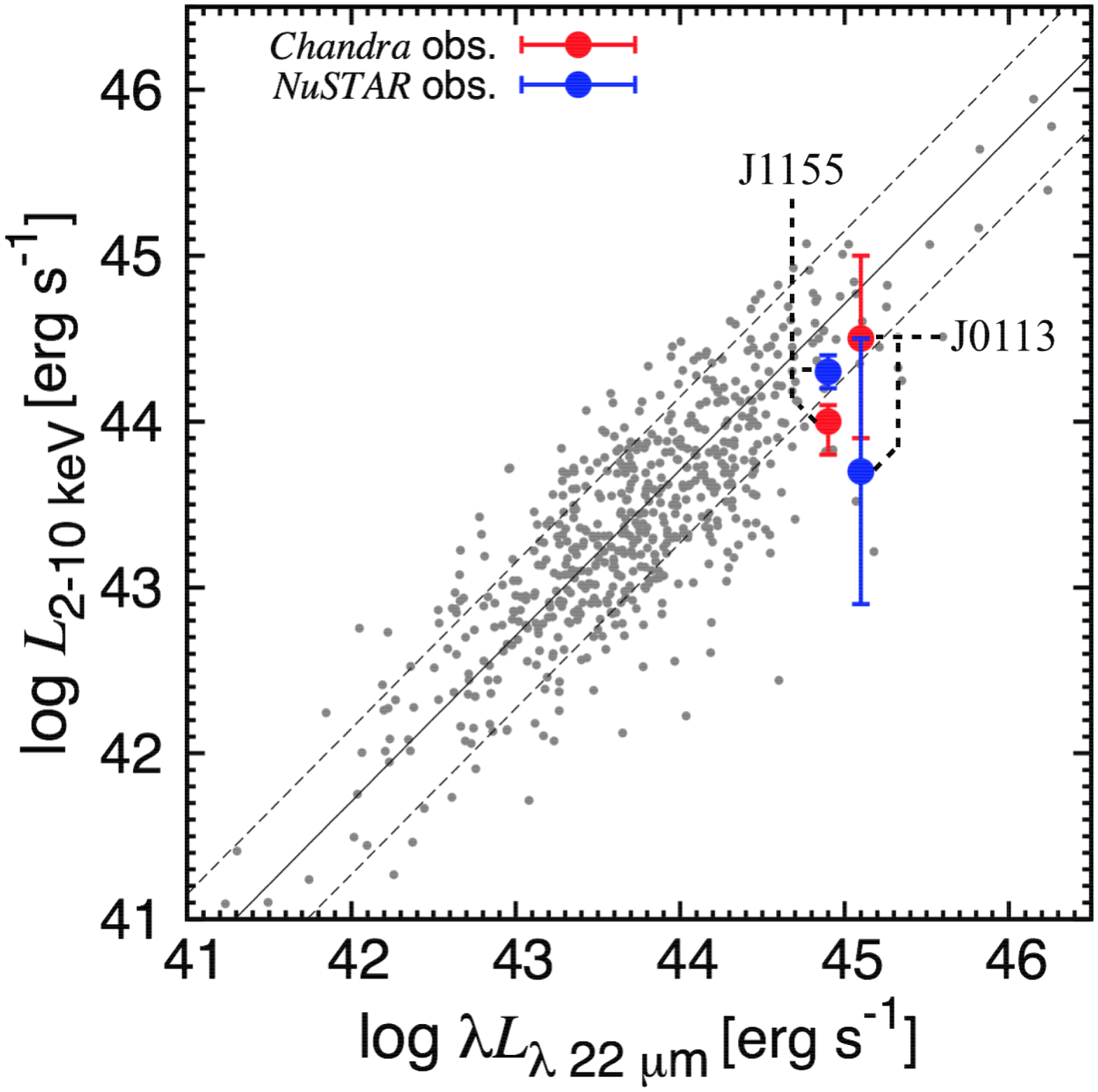}
\caption{{\label{fig:cor} 
    (Left) Correlation between the 2--10 keV and 12 $\mu$m luminosities. The solid 
    line represents the regression line obtained from the \protect\cite{Ich17} sample (gray 
    points), and the dashed lines the 1$\sigma$ envelope. The \nustar and \chandra data for
    our two LABs are shown by blue and red symbols, respectively, with 90\,\% confidence
    error bars. The arrows illustrate the time evolution for a variable AGN. The
      vertical error indicates a sudden quenching by two orders of magnitude in X-ray
      luminosity. Subsequently, the 12 $\mu$m dust emission starts to respond, with indications
      of the 20 and 40 year marks after the shutdown as estimated by \citet{Ich17b}. 
      More details about the time evolution is given in Section~\ref{sec:corMIR}.
    (Right) The same as the left figure but for the 22 $\mu$m luminosity (without time
    evolution). 
}}\vspace{.2cm}
\end{center}
\end{figure*}

\subsection{Using scaling relations to test for long-term variability}
Two possible explanations for the ionization deficits in LABs are deeply buried AGN, 
and time variable (faded) AGN (see Section \ref{sec:int}). Our previous analysis of
  individual sources in \citet{Sch16} was limited by poor knowledge of their X-ray
  obscurations. We could argue in favor of faded AGN for the sample as a whole, but not for
  individual sources. This limitation is overcome with hard X-ray ($>$ 10 keV)
  data.

  Currently, powerful ($L_{\rm X}\sim 10^{44}$ erg s$^{-1}$) and Compton-thin AGN exist
  in both J0113 and J1155 (Section~\ref{sec:broad_ana}). To test for recent, long-term
  variability, we compare their X-ray luminosities against other proxies of AGN power
  that retain some memory of the AGNs' past. These proxies are mid-infrared (MIR) emission
  from circum-nuclear dust, and [\ion{O}{III}] emission from the extended ionized gas.
  The built-in `response' times of these 'echo screens' are caused by light-travel time
  from the nucleus, their intrinsic physical properties, and also their orientation along
  the line-of-sight (a response from the parts located closer to us would arrive
  sooner at our telescopes). In the following, we check whether these proxies
  bear any signs for substantial variability in both AGN over the last $10^{2-4}$ years.

\subsubsection{The mid-infrared X-ray relation}\label{sec:corMIR}

Numerous studies have established a tight correlation between the MIR and X-ray
  luminosities in X-ray selected AGN; radiation from the central engine heats the dusty
  torus, which subsequently emits thermal radiation. This correlation is independent of
  AGN classifications
  \citep[type-1/2, Compton thickness; see e.g.][]{Gan09,Ich12,Ich17,Asm11,Asm15}. Rapid
  and prolonged fading of AGN, however, would move an individual AGN off the MIR X-ray
  correlation. \cite{Hoe11} simulated the MIR response of a dusty torus to a finite pulse
  from the AGN, as a function of the torus' structural parameters. 
  \citet{Ich17b} also estimated the typical cooling timescale of the dusty torus once the 
  AGN is suddenly quenched and even considered the dust heating with the energy exchange 
  by the gas. Accordingly, the thermal response of a torus may have decay times of 10--1000 years, 
  depending on its thickness, and in addition to the time lag caused by the light travel time
  from the nucleus to the sublimation radius. We have presented observational data supporting
  such thermal echoes in \cite{Sch16}, pending confirmation by accurately measuring
  the X-ray obscurations.

Figure~\ref{fig:cor} shows scatter plots of the absorption corrected 2--10 keV luminosity versus MIR 
(12 $\mu$m and 22 $\mu$m) luminosity for the hard X-ray selected AGN of
  \citet[][gray dots]{Ich17}. This is one of the most complete samples to date for
  the hard X-ray selected AGN. To make the plots, we converted the 14--195 keV 
  luminosities to the 2--10 keV band, using a photon index of $\Gamma=1.9$
  (i.e., $L_{\rm 14-195~keV}/L_{\rm 2-10~keV} = 2.1$). The data for J0113 and J1155 are also shown,
  using their \textit{WISE} W3 and W4 band fluxes \citep{Wri10} to estimate the 12 $\mu$m
  and 22 $\mu$m luminosities, respectively. The data for J1155 put it within the
  $1\sigma$ scattering of the reference sample. The \nustar observations of J0113 taken in
  2016, however, deviate by one order of magnitude from the best-fit correlation, whereas
  its \chandra data taken in 2014 are in much better agreement with the reference
  sample. This illustrates the effect of rapid X-ray fading on this plot. Because
  dusty tori typically have pc-scales and delayed intrinsic response times \citep{Hoe11,Ich17b},
  the MIR flux of J0113 is not expected to change much just yet (and we used the single
  WISE measurement from 2010). Sparse long-term MIR monitoring of J0113 is worth-while,
  in particular if the X-ray flux continues to fade.

  Since both AGN are still within the general scatter, we cannot infer long-term time
  ($\sim 10$--$1000$ yr) variability from the MIR X-ray relation. Note that we initially
  proposed seven targets to be observed with \nustar, covering a large range of \chandra
  fluxes. Only J0113 and J1155 were approved, because they are by far the X-ray brightest,
  maximizing chances for appreciable \nustar count rates. Therefore, highly significant
  deviations from the MIR X-ray scaling relation are not expected for these two targets,
  but could be present for the X-ray fainter LABs in our sample.

  For clarity, we illustrate the simplified time evolution (arrows) after AGN
  quenching in the left panel of Figure~\ref{fig:cor}. For this, we adopt an instantaneous
  decline by two orders of magnitude, supported by observational and theoretical data
  alike \citep{Sch10,Nov11,Sch16}. The X-ray luminosity before the quenching is set to
  $\log L_{\rm 2-10~keV} = 44.5$, the maximum luminosity we observe for J0113 and J1155. The
  12 $\mu$m luminosity is estimated using the third equation in Table~3 of
  \citet{Ich17} for luminous AGN ($43 < L_{\rm 14-195~keV} < 46$). Applying eq. (21)
  by \citet{Mar04} we derive the bolometric corrections for the 2--10 keV band,
  and then use Figure~3 of \citet{Ich17b} to estimate the cooling times.
  Note that \cite{Ich17b} assume a sudden and complete AGN shut down, whereas in reality
  it could take the AGN much longer to move back onto the general MIR X-ray correlation
  \citep[depending on the radial dependence of the dust density][]{Hoe11}.

\subsubsection{Negligible effect from MIR emission lines}

  Particularly the W3 band, covering $\sim$7.5--17
  $\mu$m\footnote{{\tt http://www.astro.ucla.edu/\textasciitilde wright/WISE/passbands.html}},
  contains several emission lines (e.g., polyclyclic aromatic hydrocarbon, [\ion{Si}{IV}] 10.5 $\mu$m 
  and [\ion{Ne}{II}] 12.8 $\mu$m) 
  from the more extended ($\gtrsim$ 100 pc) star-forming and narrow-line regions than that of 
  the torus ($\lesssim$ 10 pc).   These could contribute to the 
  observed MIR flux, potentially diluting the expected MIR response from the dusty torus to a 
  change in AGN luminosity. We can probe for this effect considering the X-ray
  MIR correlations obtained using low and high spatial resolution observations:
  If the emission lines are bright, then the observed MIR flux would increase systematically
  with lower resolution, as more emission line flux is included from a larger volume around the
  AGN. However, this trend is not found when comparing the high-resolution study of
   \citet{Asm15} with the low-resolution study of \citet{Ich17} (see Figure~3 
  of \citealt{Ich17}). We conclude that the emission lines do not alter the MIR fluxes
  significantly.

  Even if the W3 band was highly contaminated, one could still use the W4 band
  ($\sim$ 20--26 $\mu$m), which is less affected. The [\ion{O}{IV}] line at
  25.89 $\mu$m, which is a bright line in AGN spectra
  \citep{Tre10}, is lower than the X-ray / MIR luminosities by two orders of
  magnitude \citep[e.g.,][]{Liu14,Kaw16}. Unless the AGN and their kpc-scale
  surroundings in our low-$z$ LABs are very different from typical AGN, MIR emission
  lines are not expected to skew the MIR X-ray scaling relations.

\subsection{The [\ion{O}{III}] X-ray relation}\label{sec:corOIII}
\begin{figure}[!ht]
\begin{center}
\includegraphics[scale=0.37]{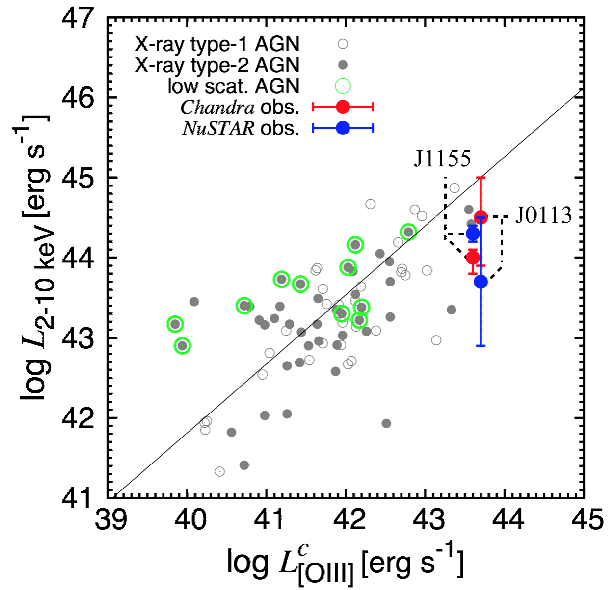}
\caption{\label{fig:cor_x2oiii} 
  Correlation between the 2--10 keV and corrected [\ion{O}{III}] luminosities. 
  For clarity, J0113 is slightly shifted to the right. The solid
  line represents the regression line obtained by \cite{Ued15}. The filled and un-filled
  circles denotes the X-ray type-1 ($\log(N_{\rm H}/$cm$^{-2}) < 22$) and X-ray type-2
  ($\log(N_{\rm H}/$cm$^{-2}) \geq 22$) AGN, respectively. The superimposed green circles 
  are attached to the low scattering AGN ($f_{\rm scat} < 0.5$\%). Objects without detection
  of the H$\beta$ line were excluded from this plot because no extinction correction
  could be obtained for them.}
\end{center}
\end{figure}

[\ion{O}{III}]$\lambda5007$ emission is another proxy commonly used to infer AGN power 
\cite[e.g.,][]{Mul94}. A typical [\ion{O}{III}] nebular size observed in local Seyfert 
galaxies can reach several kpc \citep[e.g.,][]{Sch03,Ben06}. In Figure~\ref{fig:cor_x2oiii}, 
we compare our LABs with local AGN detected in the {\it Swift}/BAT 9-month catalog \citep{Ued15}. 
Both LABs have lower X-ray luminosities at a given [\ion{O}{III}] luminosity than predicted
from the average properties of the {\it Swift}/BAT sample, yet not at a level that would
distinguish them as fading. The apparently high [\ion{O}{III}] luminosity could simply be a fact
of our selection (targetting optically over-luminous nebulae), and that our type of targets
is not included in the commonly used SDSS AGN catalogs
\citep[the latter are incomplete concerning extreme sources, see][]{Sch13,Bar17}.
Nonetheless, for J0113 we know that it has been fading for the last two years. In addition,
\cite{Ued15} report a lower [\ion{O}{III}] to X-ray luminosity ratio on average for low scattering 
fraction AGN ($f_{\rm scat} < 0.5$\%; green circles in Figure~\ref{fig:cor_x2oiii}), to which
J0113 belongs. J0113 is 10--50 times more luminous in [\ion{O}{III}] than other low
scattering AGN. This could simply be a consequence of poor statistics at the high luminosity
end, or a consequence of variability (the [\ion{O}{III}] X-ray relation decorrelates more
easily than the MIR relation under AGN flickering due to the slower [\ion{O}{III}] response).
We conclude that, for our particular two targets, the [\ion{O}{III}] X-ray relation does not bear
conclusive evidence for AGN flickering.




\section{Summary}\label{sec:sum}

Using 20 ksec \nustar observations, we probed the hard X-ray emission of two recently
discovered \citep{Sch16} low-$z$ LABs, J0113 and J1155. We detect them in the observed 
3--30 keV band on the $8\sigma$ and $21\sigma$ level, respectively. Adding \chandra
archival data, we performed a broadband X-ray spectral analysis of the 0.5--30 keV band.
The spectra were well reproduced with a base-line model consisting of an absorbed cut-off 
power-law, reflection components from distant, cold matter, and scattered emission.
This is the most detailed X-ray analysis of AGN in any LABs published to date. It also
shows that with \nustar one can strongly constrain the primary X-ray luminosity. 

For J1155, we find a moderate scattering fraction of $f_{\rm scat} = 1.2$\,\%. This shows 
that the AGN is not deeply buried in a torus, whose existence is suggested by the detection 
of the iron-K$\alpha$ line. The obscuration is moderate with
$\log\,(N_{\rm H}/$cm$^{-2}) \sim 22.8$. Hence, the AGN emission can indeed ionize the
narrow line region that extends out to a radius of $\gtrsim30$\,kpc, over a wide opening
angle. AGN ionization of the gas is required by its high
${\rm log}($[\ion{O}{III}]/H$\beta)$ line ratio \citep{Sch13}.

Regarding J0113, the torus may be highly developed. It reduces the scattering gas as inferred 
from the undetected scattered emission (upper limit $f_{\rm scat} < 9.9$\,\%) as well as 
from the high column density ($\log\,(N_{\rm H}/$cm$^{-2}) \sim 23.7$). This is consistent with 
the collimated bipolar outflow seen in the optical [\ion{O}{III}] emission. However, the 
central $2$--$4$ kpc also have high [\ion{O}{III}] surface brightness, showing that ionizing
radiation can leak out in other directions as well. 

Our optical 3D spectroscopy with Gemini/GMOS
\citep[and imaging data from our previous work by][]{Sch16} reveals two very different
systems. For J0113 we find multiple evidence for recurrent powerful outflow events
over several 10 million years. The prominent bipolar outflow in J0113 is accelerating,
as gas velocities increase with distance from the nucleus. This behaviour is best explained
by AGN flickering \citep{Zub16}. J0113 is a prototypical example of the outflow formation
process suggested by \cite{Fau12}, where hot buoyant gas breaks out along paths of least
resistance, inflating kpc-scale superbubbles. The extended narrow-line region around
J1155, on the other hand, reveals a very different morphology. It is more consistent
with an advanced stage, where the remnants of multiple outflows are dispersing in the
halo, passively illuminated/ionized by a current powerful accretion event.
Deeper 3D spectra taken over a larger area, and covering more diagnostic emission lines,
would allow us to better constrain the gas physics and formation histories of these two
distinguished targets.

Common to both systems are low velocity dispersions of $100$--$150$
km s$^{-1}$. In particular, the AGN themselves are not detectable in the line
width maps. In case of J0113, severe dust extinction could be hiding more turbulent
gas closer to the nucleus. Also, previous outflows could have cleared the escape
paths for the current streams, reducing mechanical resistance and thus turbulence.

In terms of [\ion{O}{III}] extent and luminosity, we find our low-$z$ LABs
indistinguishable from high-$z$ LABs. J1155, in particular, resembles one of the
best studied LABs \citep[B1, see e.g.,][]{Fra01,Pal04,Ove13} in many other respects,
hosting a powerful AGN.

J0113 and J1155 fall within the intrinsic scatter of the MIR X-ray luminosity
correlation, as well as within the [\ion{O}{III}]-to-X-ray relation of the hard X-ray
selected AGN \citep[][]{Ich17}. J0113 has faded by a factor $\sim$ 5 over two years
of time between our \chandra and \nustar observations, but not enough to distinguish
it with respect to the other AGN in these relations. It is worthwhile to monitor the
MIR emission of J0113, roughly on a yearly basis, in particular if the X-ray luminosity
continues fading. This would bear clues about the size and structure of its obscuring
dusty torus, and the effect of long-term AGN variability on the intrinsic scatter of the
MIR X-ray relation.

Our targets are among the most luminous [\ion{O}{III}] emitters in 
the Universe, and at $z\sim0.3$ they are easy targets for follow-up studies.\\


\acknowledgments
This work was financially supported by the Grant-in-Aid for JSPS fellows for young
researchers (TK) and also for Scientific Research 40756293 (KI). Further financial
support was provided by the Gemini Observatory (TK). MS acknowledges support by the
National Aeronautics and Space Administration through \chandra Award Number
GO4-15110X issued by the \chandra X-ray Observatory Center, which is
operated by the Smithsonian Astrophysical Observatory for and on behalf of the National
Aeronautics Space Administration under contract NAS8-03060. This research has made use
of the NASA/IPAC Infrared Science Archive, which is operated by the Jet Propulsion
Laboratory, California Institute of Technology, under contract with the National
Aeronautics and Space Administration.

Author contributions: TK reduced the \nustar and \chandra X-ray data, and performed the
joint spectral analysis; TK and MS wrote the manuscript; MS performed the analysis of the
3D spectra, and led the \nustar, \chandra and Gemini proposals; JT reduced the Gemini 3D
spectra and obtained the [\ion{O}{III}] luminosities; RD obtained the velocity, velocity
dispersion and line flux maps.

MS thanks Claudio Ricci, Daniel Asmus and Ai-Lei Sun for sharing their expertise on the
subject.

This work is based on data from the \nustar mission, a project led by the California Institute of Technology, managed by the Jet Propulsion Laboratory, and funded by NASA. This research has made use of the \nustar Data Analysis Software (NuSTARDAS) jointly developed by the ASI Science Data Center and the California Institute of Technology.

This work made use of data supplied by the UK Swift Science Data Centre at the University of Leicester.

Based on observations made by the \textit{Chandra} X-ray Observatory, and on data obtained from the \textit{Chandra} Data Archive. We also made use of the software provided by the \textit{Chandra} X-ray Center (CXC) in the application packages CIAO. 

Based on observations obtained at the Gemini Observatory, which is operated by the
Association of Universities for Research in Astronomy, Inc., under a cooperative agreement
with the NSF on behalf of the Gemini partnership: the National Science Foundation (United
States), the National Research Council (Canada), CONICYT (Chile), Ministerio de Ciencia,
Tecnolog\'{i}a e Innovaci\'{o}n Productiva (Argentina), and Minist\'{e}rio da Ci\^{e}ncia,
Tecnologia e Inova\c{c}\~{a}o (Brazil). 



\end{document}